\documentclass[lettersize,journal]{IEEEtran}
\usepackage{amsmath,amsfonts}
\usepackage{algorithmic}
\usepackage{algorithm}
\usepackage{array}
\usepackage[caption=false,font=normalsize,labelfont=sf,textfont=sf]{subfig}
\usepackage{textcomp}
\usepackage{stfloats}
\usepackage{url}
\usepackage{verbatim}
\usepackage{graphicx}
\usepackage{cite}
\usepackage{multirow}
\usepackage{makecell}
\usepackage{booktabs}
\usepackage{xcolor}

\newif\ifrevision
\revisionfalse    

\ifrevision
    \newcommand{\rev}[1]{\textcolor{blue}{#1}}
\else
    \newcommand{\rev}[1]{#1}
\fi

\begin{document}

\title{Adaptive Rank Allocation for Federated Parameter-\\Efficient Fine-Tuning of Language Models}

\author{
    Fei~Wu,
    Jia~Hu,
    Geyong~Min,
    and~Shiqiang~Wang,~\IEEEmembership{Fellow,~IEEE}
    \thanks{This work was supported in part by UK Engineering and Physical Sciences Research Council (EPSRC) Grant No. EP/X019160/1, Research and Innovation (UKRI) Grant No. EP/Y036786/1, and Horizon Europe Grant No. 101129910. For the purpose of open access, the author has applied a Creative Commons Attribution (CC BY) licence to any Author Accepted Manuscript version arising. \textit{(Corresponding authors: Jia Hu and Geyong Min.)} }
    \thanks{Fei Wu, Jia Hu, Geyong Min and Shiqiang Wang are with the Department of Computer Science, University of Exeter, EX4 4PY Exeter, UK E-mail: \{fw407, j.hu, g.min, s.wang9\}@exeter.ac.uk.}%
}

\markboth{Journal of \LaTeX\ Class Files,~Vol.~X, No.~X, XXXX~2025}%
{Shell \MakeLowercase{\textit{et al.}}: A Sample Article Using IEEEtran.cls for IEEE Journals}


\maketitle

\begin{abstract}
Pre-trained Language Models (PLMs) have demonstrated their superiority and versatility in modern Natural Language Processing (NLP), effectively adapting to various downstream tasks through further fine-tuning. Federated Parameter-Efficient Fine-Tuning (FedPEFT) has emerged as a promising solution to address privacy and efficiency challenges in distributed training for PLMs on resource-constrained local devices. However, our measurements reveal two key limitations of FedPEFT: heterogeneous data across devices exacerbates performance degradation of low-rank adaptation, and a fixed parameter configuration results in communication inefficiency.
To overcome these limitations, we propose FedARA, a novel adaptive rank allocation framework for federated parameter-efficient fine-tuning of language models. 
Specifically, FedARA employs truncated Singular Value Decomposition (SVD) adaptation to enhance similar feature representation across clients, significantly mitigating the adverse effects of data heterogeneity. 
Subsequently, it utilizes dynamic rank allocation to progressively identify critical ranks, effectively improving communication efficiency. 
Lastly, it leverages rank-based module pruning to automatically remove inactive modules, steadily reducing local computational cost and memory usage in each federated learning round. 
Extensive experiments show that FedARA consistently outperforms baselines by an average of 6.95\% to 8.49\% across various datasets and models under heterogeneous data while significantly improving communication efficiency by 2.40\(\times\). Moreover, experiments on various edge devices demonstrate substantial decreases in total training time and energy consumption by up to 48.90\% and 46.95\%, respectively.
\end{abstract}

\begin{IEEEkeywords}
Federated Learning, Communication Efficiency, Data Heterogeneity, Parameter-Efficient Fine-Tuning.
\end{IEEEkeywords}

\section{Introduction}
\IEEEPARstart{P}{re-trained} Language Models (PLMs) \cite{vaswani2017attention, sanh2019distilbert, devlin2019bert, lewis2019bart} have demonstrated extraordinary performance and foundational capabilities in modern Natural Language Processing (NLP). Fine-tuning of PLMs further enhances their performance on downstream tasks, such as sequence classification, summarization, and question answering. Pre-training and fine-tuning are evolving into two distinct yet mutually reinforcing stages in NLP. 
1) \textbf{Pre-training:} it involves self-supervised training on large-scale corpora from sources like books, news articles, and Wikipedia, aiming to learn general representations of contextual language and create a generalizable pre-trained model.
2) \textbf{Fine-tuning:} it extends the pre-trained model to specific downstream tasks through supervised training, optimizing its performance to meet domain requirements.
Overall, these two stages greatly enhance the model's generality and flexibility across diverse NLP tasks.

Although PLMs have demonstrated superior performance on general-purpose tasks \cite{brown2020language, kang2024toex}, they still require fine-tuning with local data to achieve optimal results in specific domains. However, since this data is typically generated by users, directly collecting it poses a significant risk of privacy leakage. To address this issue, Federated Learning (FL) is considered an effective approach \cite{mcmahan2017communication}, as it protects data privacy by keeping raw data on local devices, ensuring compliance with regulations such as \textit{GDPR} \cite{GDPR} and the \textit{Data Protection Act} \cite{DPA2018}. The FL process involves a central server orchestrating the aggregation of local parameter updates from clients.

\begin{figure}[!t]
  \centering
  \includegraphics[width=1.0\linewidth]{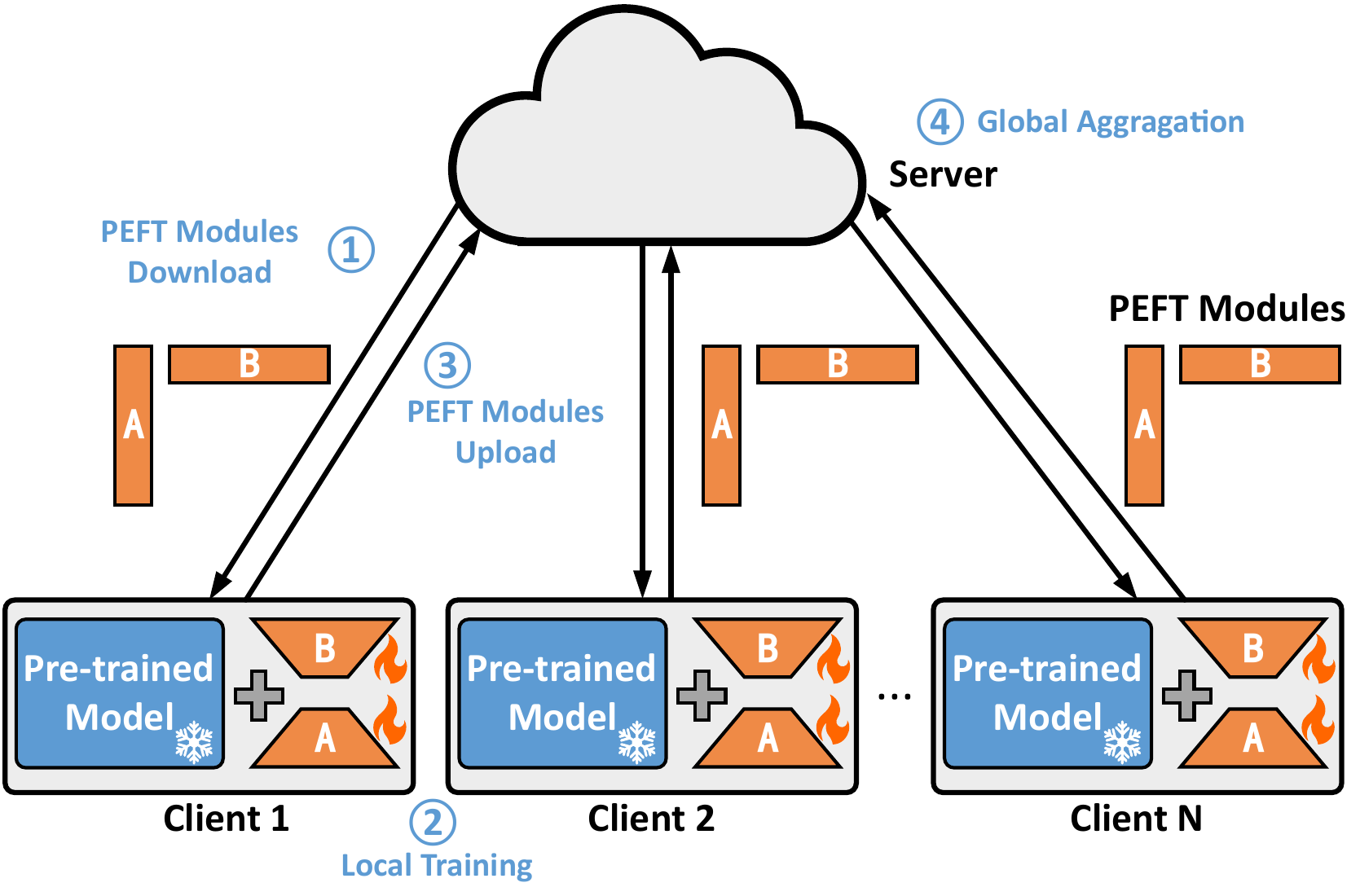}
  \caption{The complete training process of Federated Parameter-Efficient Fine-Tuning (FedPEFT), where local updates utilize Low-Rank Adaptation (LoRA), while keeping the pre-trained model frozen.}
  \label{fig1}
\end{figure} 

However, the massive scale of PLMs, often with hundreds of millions to billions of parameters~\cite{dubey2024llama}, incurs substantial computation and communication costs in FL, rendering deployment on resource-constrained devices impossible.
To address this issue, a promising solution is to adopt Parameter-Efficient Fine-Tuning (PEFT) methods, which train only a minimal set of parameters using adaptable modules while keeping the pre-trained model frozen \cite{houlsby2019parameter, pfeiffer2021adapterfusion, lester2021power, li2021prefix, hu2022lora}. Among these PEFT methods, Low-Rank Adaptation (LoRA) \cite{hu2022lora} has gained widespread adoption for its efficiency, achieved by introducing two trainable low-rank matrices for diverse task adaptation. By combining PEFT with FL, distributed data resources on local devices can be fully leveraged for collaborative fine-tuning in a privacy-preserving manner. Consequently, Federated Parameter-Efficient Fine-Tuning (FedPEFT) shows significant advancements and finds wide-ranging applications across various domains \cite{zhang2024towards, wang2024flora, bai2024federated, babakniya2023slora, yan2024federa, yang2024sa, sun2024improving}. Building on this progress, FedPEFT has increasingly gained recognition as a key method for distributed fine-tuning on local devices \cite{cai2023efficient, Woisetschlager2024federated}. Figure~\ref{fig1} shows the complete FedPEFT training process.

Nevertheless, FedPEFT still faces \textit{two major challenges} when applied to local devices. 
\rev{\textbf{1) FedPEFT may exacerbate the impact of non-IID data on model performance.} 
Performance degradation caused by non-independent and identically distributed (non-IID) data, where different clients hold data from distinct distributions, is a well-known challenge in FL.} Our preliminary measurements show that FedPEFT exacerbates this issue. Specifically, on the \textit{20News} dataset \cite{lang1995newsweeder} using the \textit{DistilBERT} model~\cite{sanh2019distilbert}, FedPEFT methods exhibit an average accuracy drop of 14.62\% from non-IID to IID settings---more than double the 6\% drop observed with federated full fine-tuning \cite{babakniya2023slora}. 
\rev{\textbf{2) Fixed-rank configurations in FedPEFT may incur dual losses in performance and communication cost compared with dynamic-rank allocations.} 
These dual losses arise not because rank choices inherently harm accuracy, but rather because static allocations cannot accommodate the heterogeneous sensitivity of different modules under a constrained total rank budget.}
In particular, static allocations often underfit high-sensitivity positions, leading to suboptimal performance---our experiments reveal accuracy gaps of up to 9.76\% across different configurations. Meanwhile, redundant parameters assigned to less critical components are repeatedly transmitted during FL rounds without corresponding performance gains, resulting in communication inefficiency.

These losses become more severe under non-IID settings, where data heterogeneity intensifies the subspace divergence of federated low-rank adaptation among client-specific models. This divergence leads to greater variability in rank requirements, making fixed-rank configurations increasingly misaligned with actual capacity needs. Consequently, both performance degradation and communication overhead are further amplified.  
Therefore, these two challenges should be jointly addressed to fully unlock the potential of FedPEFT in practical FL scenarios.

To address these challenges, we propose \textbf{A}daptive \textbf{R}ank \textbf{A}llocation for \textbf{Fed}erated PEFT of language models (\textbf{FedARA}), designed to enhance performance and communication efficiency with minimized resource costs by integrating \textit{three novel methods:}
\textbf{1) Truncated SVD adaptation (Section \ref{method1}).} 
We hypothesize that the performance degradation of LoRA arises from its inherent structure and asymmetric initialization. Inspired by Singular Value Decomposition (SVD), we propose a novel truncated SVD adaptation scheme that inserts a diagonal matrix into the low-rank subspace to enable independent feature scaling while preserving symmetric initialization.  This mitigates client drift and enhances cross-client feature alignment, significantly alleviating the negative impact of non-IID data.
\textbf{2) Dynamic rank allocation (Section \ref{method2}).} 
To avoid fixed parameter configurations, we present a dynamic rank allocation method that generates local rank masks through triplets on each client and produces global rank masks via threshold-based arbitration on the server. During federated fine-tuning, less critical ranks are progressively pruned, retaining only significant ones based on global rank masks, effectively improving the communication efficiency of FedPEFT.
\textbf{3) Rank-based module pruning (Section \ref{method3}).} 
Building on the above two methods, the rank of SVD modules dynamically evolves over FL rounds, we thereby adopt a rank-based module pruning mechanism to automatically remove inactive modules during training. By continuously monitoring rank changes, modules with ranks that fall to zero are safely excluded from the trainable parameters, steadily reducing the computational and storage overhead for client-side training.

We implemented FedARA and evaluated it on three representative edge devices, NVIDIA Jetson AGX Orin, Orin Nano, and Raspberry Pi 5, whose hardware profiles reflect typical resource-constrained settings. We utilize three pre-trained models across five diverse NLP datasets. FedARA outperforms baselines \cite{sun2024improving, zhang2023lora, babakniya2023slora, yan2024federa} on classification tasks by an average of 6.95\% to 8.49\% under non-IID data while significantly improving communication efficiency by an average of 2.40\(\times\). Remarkably, the total training time on the above devices also decreases by 27.65\%, 46.57\%, and 48.90\%, respectively, and the energy consumption drops by 46.95\% on Orin Nano. Ablation experiments reveal that truncated SVD adaptation improves the average accuracy by 7.71\%, dynamic rank allocation reduces communication overhead by 70.82\% per round, and rank-based module pruning lowers the average local computation and peak GPU memory usage per round by 10.81\% and 31.67\%, respectively.

We summarize our key contributions below:
\begin{itemize}
    \item We propose FedARA, the first comprehensive approach to address both the adverse effects of non-IID data and fixed parameter configurations in FedPEFT for resource-constrained devices.
    \item We create and integrate three innovative methods into FedARA: 
    1) truncated SVD adaptation strengthens the structure and adjusts initialization to enhance similar feature representation across clients;
    2) dynamic rank allocation generates rank masks locally and arbitrates them globally to reduce the communication of less important parameters;
    3) rank-based module pruning monitors dynamic rank changes to remove inactive modules.
    \item We validate the effectiveness of FedARA through extensive experiments across various datasets and models on advanced edge devices, demonstrating that FedARA consistently improves performance under non-IID data and significantly boosts system efficiency in both communication and computation.
\end{itemize}

The rest of the paper is organized as follows:
Section~\ref{background} introduces the background of FedPEFT and explains the motivation of our work.
Section~\ref{overview} presents the proposed FedARA in terms of system architecture and algorithm process. 
Section~\ref{design} details the three innovative methods in FedARA.
Section~\ref{setup} describes the complete experimental setup.
Section~\ref{evaluation} shows and analyzes the experimental results, including ablation studies and extended experiments.
Section~\ref{relatedwork} reviews the related work.
Finally, Section~\ref{conclusion} concludes the paper.

\section{Background and Motivations}
\label{background}

\subsection{Federated Parameter-Efficient Fine-Tuning (FedPEFT)} \label{fedpeft}
\begin{figure*}[t]
  \centering
  \subfloat[FedFFT vs. FedPEFT]{%
    \includegraphics[width=0.24\textwidth]{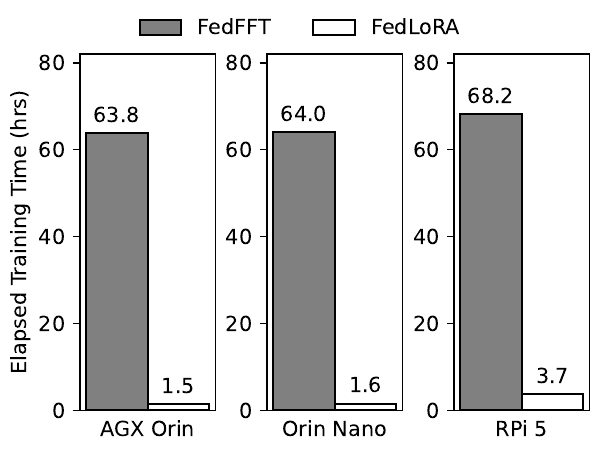}%
    \label{fig2-a}}
  \hfill
  \subfloat[The effect of non-IID]{%
    \includegraphics[width=0.24\textwidth]{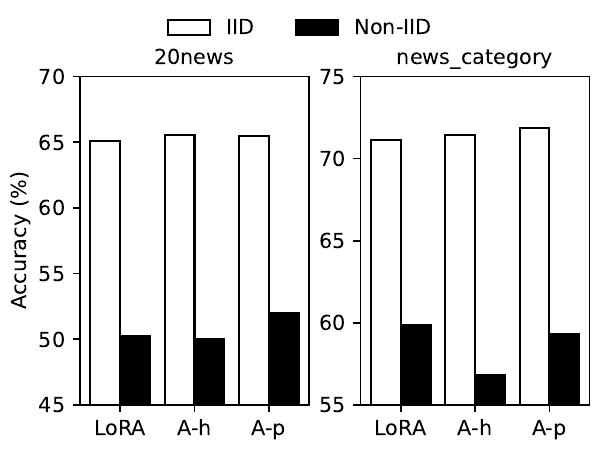}%
    \label{fig2-b}}
  \hfill
  \subfloat[The impact of positions]{%
    \includegraphics[width=0.24\textwidth]{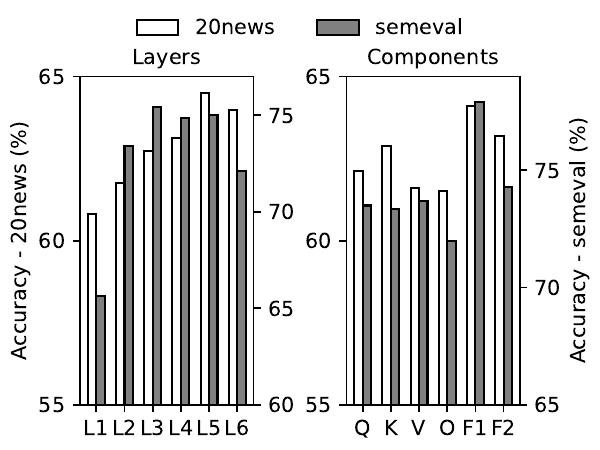}%
    \label{fig2-c}}
  \hfill
  \subfloat[FedPEFT bottlenecks]{%
    \includegraphics[width=0.24\textwidth]{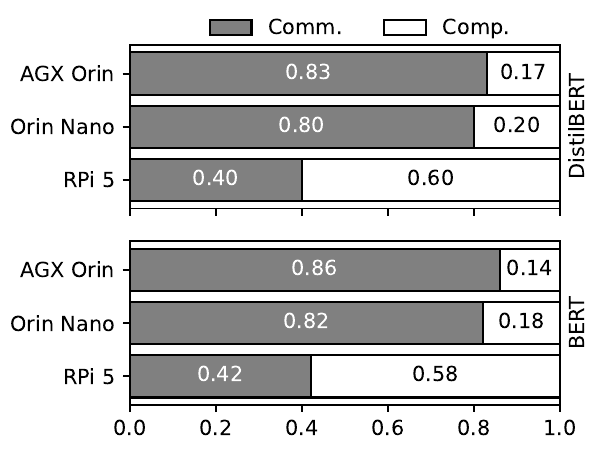}%
    \label{fig2-d}}
  \caption{The preliminary measurement results of FedPEFT in a homogeneous FL setting: (a) Comparison of overall training time on different edge devices (\textit{BERT}; \textit{20News}; Rank=12; Batch size=4). (b) Accuracy of FedPEFT under IID and non-IID scenarios (\textit{DistilBERT}; Rank=24; Adapter-h (A-h) size=108; Adapter-p (A-p) size=216). (c) Impact of inserted positions on accuracy (\textit{DistilBERT}; Rank=24). (d) Communication and computation time percentage across models and devices using FedLoRA (\textit{News Category}; Rank=12; Batch size=4).}
  \label{fig:main}
\end{figure*}

FedPEFT orchestrates PEFT within the FL framework to efficiently update local models, drastically reducing communication overhead by only transmitting a small number of PEFT modules. These PEFT modules can be replaced by Adapter \cite{houlsby2019parameter, pfeiffer2021adapterfusion}, Prompt\cite{lester2021power, li2021prefix}, or LoRA \cite{hu2022lora}. Among them, LoRA-based methods are especially popular because they introduce no additional latency during inference. FedPEFT not only preserves privacy but also effectively aggregates diverse knowledge across clients, facilitating the collaborative development of a high-quality global model. By harnessing the advantages of both PEFT and FL, FedPEFT has become an essential technique for deploying fine-tuning on resource-constrained devices \cite{cai2023efficient, Woisetschlager2024federated, yan2024federa, bai2024federated, wang2024flora, yang2024sa}.

As illustrated in Figure \ref{fig1}, FedPEFT training process unfolds as follows.
\textbf{Step 1:} A cloud server selects a subset of clients to broadcast the initial parameters.
\textbf{Step 2:} On these clients, the weights of the pre-trained model are frozen, and trainable PEFT modules are inserted for local training.
 \textbf{Step 3:} Only these PEFT modules are updated to the cloud server from each client.
\textbf{Step 4:} The cloud server aggregates these updates from different clients.
This cycle (Steps 1-4) is repeated for multiple rounds in FL until the model converges.

\subsection{Preliminary Measurements} \label{pre-experiments}
We conducted a series of preliminary measurements to explain the motivations behind the design of FedARA.

\textbf{Observation 1: FedPEFT is more effective than FedFFT on advanced edge devices.} 
FedPEFT comprises two main methods: adapter-based \cite{cai2023efficient} and LoRA-based methods \cite{zhang2024towards, wang2024flora, bai2024federated, babakniya2023slora, yang2024sa}. LoRA-based methods have gradually replaced adapter-based methods to avoid additional inference latency.
According to established work \cite{cai2023efficient}, FedPEFT achieves comparable accuracy to federated full fine-tuning (FedFFT) while significantly reducing local training time by 1.6$\times$, communication cost by 200$\times$, and memory usage by 1.5$\times$ on previous-generation edge devices like NVIDIA Jetson TX2, Nano, and Raspberry Pi 4B. 
However, its impact on overall training time on advanced devices remains underexplored. To fill this gap, we evaluate FedLoRA on the \textit{20News} \cite{lang1995newsweeder} dataset with the \textit{BERT} \cite{devlin2019bert} model using the latest edge devices. Following FedPEFT's typical settings \cite{cai2023efficient}, we set the communication bandwidth to 1MB/s. As depicted in Figure \ref{fig2-a}, the results indicate that FedFFT's overall training time is 42\(\times\), 40\(\times\), and 18\(\times\) longer than FedLoRA on AGX Orin, Orin Nano, and RPi 5, respectively. FedPEFT demonstrates its effectiveness compared to FedFFT for cutting-edge embedded devices.

\textbf{Observation 2: The performance of FedPEFT significantly decreases under non-IID data.} Figure \ref{fig2-b} illustrates the accuracy of different FedPEFT methods using the \mbox{\textit{DistilBERT}} \cite{sanh2019distilbert} model for both IID (white bars) and non-IID data (black bars). The non-IID partitioning follows the FedAvg setting \cite{mcmahan2017communication}, with a severe class imbalance in data distribution. The parameter sizes for Adapter-h (A-h) \cite{houlsby2019parameter} and Adapter-p (A-p) \cite{pfeiffer2021adapterfusion} are set to match those of LoRA. Our measurements indicate a significant performance drop in all FedPEFT methods under non-IID conditions compared to IID conditions, with average accuracy reductions of 14.62\% and 12.81\% on the \textit{20News} \cite{lang1995newsweeder} and \textit{News Category} \cite{misra2022news} datasets, respectively. Existing research lacks effective strategies to mitigate the adverse impact of non-IID data on performance.

\textbf{Observation 3: The contribution of PEFT modules in FedPEFT to performance varies with their inserted positions, leading to suboptimal performance and inefficient communication in fixed parameter configurations.}  
In FedPEFT, low-rank modules can be placed at various layers and components, such as \textit{Query (Q)}, \textit{Key (K)}, \textit{Value (V)}, \textit{Out (O)}, \textit{F1}, and \textit{F2}. Here, \textit{O} denotes the linear layer in the multi-head attention block, while \textit{F1} and \textit{F2} represent the two linear layers of the feed-forward network. As shown in Figure \ref{fig2-c}, we allocated a fixed rank budget to specific components or layers to evaluate their impact on accuracy. The left sub-figure shows the allocation of the entire rank budget to a single layer, while the right sub-figure illustrates its distribution across a single component in all layers. The findings reveal that deeper layers near the output often have a more significant effect than shallower ones, with components \textit{F1} and \textit{F2} exerting a greater influence than component \textit{O}. Moreover, the impact of these inserted positions varies across datasets, suggesting that there is no one-size-fits-all configuration. Fixed-rank configurations result in inefficient communication, as they repeatedly transmit less impactful parameters over rounds, yielding minimal performance gains under a limited parameter budget.

\textbf{Observation 4: Bottlenecks differ across different types of advanced edge devices.} In experiments with LoRA-based methods \cite{zhang2024towards, wang2024flora} on \textit{News Category} \cite{misra2022news}, given a communication bandwidth of 1MB/s, we found that the bottlenecks varied depending on the device type, as shown in Figure \ref{fig2-d}. On high-end devices, communication is the predominant bottleneck, with communication-to-computation ratios of 4.44 and 5.35 for different models. In contrast, on Raspberry Pi devices, computation becomes the primary bottleneck, with communication-to-computation ratios of 0.67 and 0.72. Importantly, the bottleneck shifts towards communication as model parameters increase, consistent with the recent study \cite{Woisetschlager2024federated}.

\begin{figure}[!t]
  \centering
  \includegraphics[width=0.9\linewidth]{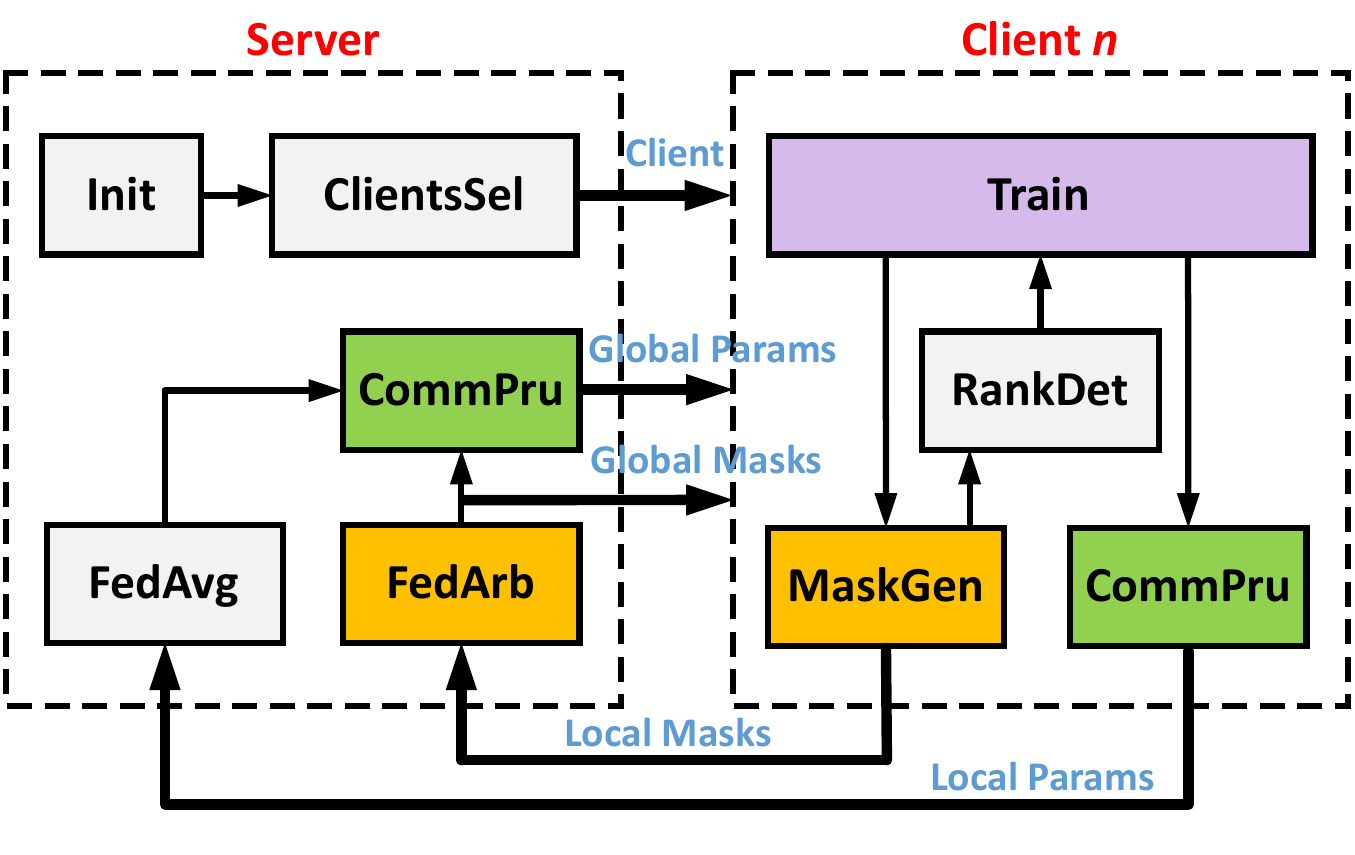}
  \caption{System architecture of FedARA, illustrating the interaction between the server and client, including ClientsSel (Client Selection), CommPru (Communication Pruning), FedArb (Federated Arbitration), MaskGen (Rank Mask Generation), and RankDet (Local Rank Detection).}
  \label{fig3}
\end{figure}

\textbf{Insights:} These observations underscore key directions for advancing our research. 
First, given FedLoRA's advantages, such as improved training efficiency, future research could benefit from moving beyond comparisons with FedFFT and instead focusing on addressing the unique challenges it presents in FL.
Second, the performance degradation of different PEFT methods under non-IID remains a critical barrier, necessitating further investigation into effective approaches. 
Third, fixed parameter configurations lead to communication inefficiency by repeatedly transmitting less impactful parameters over multiple FL rounds, particularly in resource-constrained environments. 
Finally, optimizing FedPEFT for edge devices requires further refinement, as different hardware presents distinct bottlenecks.
To bridge these gaps, enhancing fine-tuning structures and dynamically adjusting parameters could be essential while simultaneously minimizing both communication and computational overhead.

\section{Overview of FedARA}
\label{overview}
Due to LoRA's limitations in handling complex data heterogeneity and the communication inefficiency of fixed parameter configurations, we propose FedARA, an adaptive rank allocation framework for FedPEFT of language models. Our approach introduces a truncated SVD structure to better address data heterogeneity in federated settings. Meanwhile, we incorporate a dynamic rank allocation algorithm to enhance communication efficiency. Additionally, we employ rank-based module pruning to reduce computational time and peak memory usage per round in clients.

The system architecture of the FedARA framework, shown in Figure~\ref{fig3}, includes a server and multiple clients. Clients utilize the truncated SVD adaptation to train local models and generate local masks, while the server aggregates the global model and arbitrates global masks. The ranks of the global model are dynamically allocated through the arbitrated global masks. Communication between the server and clients involves transmitting masks and progressively pruned parameters. Local clients detect rank changes in each round to remove inactive SVD modules safely.

\begin{algorithm}[t]
\caption{Federated Adaptive Rank Allocation (FedARA)}
\label{alg1}
\begin{algorithmic}[1]
\STATE \textbf{Input:} all clients $C$, rounds $R$, initial SVD modules $P_0$, initial global rank masks $M_0$, threshold $T_h$, target average rank $T_r$, test dataset $D_t$, local datasets $D_l$, the number of trainable parameters $N_t$
\STATE \textbf{Output:} optimized SVD modules $P^*$, optimized global rank masks $M^*$
\STATE {\text{SERVER:}}
\STATE Initialize $P_g \gets P_0$, $M_g \gets M_0$
\FOR{$r \gets 1$ to $R$}
    \STATE $C_s \gets \textbf{ClientsSel}(C)$
    \STATE // Prune broadcasting parameters by $M_g$:
    \STATE $\hat{P}_g \gets \textbf{CommPru}(P_g, M_g)$
    \STATE $\textbf{BroadcastParams}(C_s, \hat{P}_g, M_g)$
    \STATE // Parallelism among different clients:
    \STATE $\{\hat{P}_{l+1}\}, \{M_{l+1}\} \gets \text{CLIENT\_TRAIN}(r, \hat{P}_g, M_g)$
    \STATE // Aggregation based on FedAvg:
    \STATE $P_{g+1} \gets \textbf{FedAvg}(\{\hat{P}_{l+1}\}, M_g)$
    \STATE // Arbitration based on the threshold $T_h$:
    \STATE $M_{g+1} \gets \textbf{FedArb}(\{M_{l+1}\}, T_h, M_g)$
\ENDFOR
\STATE $P^* \gets P_g, M^* \gets M_g$
\STATE Evaluate final model: $acc \gets \textbf{Eval}(P^*, D_t)$
\STATE \textbf{return} $acc, P^*, M^*$
\STATE {\text{CLIENT\_TRAIN}}$(r, \hat{P}_g, M_g)$
\STATE \hspace{0.5cm} $P_l \gets \hat{P}_g$, $M_l \gets M_g$
\STATE \hspace{0.5cm} $P_{l+1} \gets \textbf{Train}(D_l ,P_l, N_t)$
\STATE \hspace{0.5cm} // Generate local rank masks:
\STATE \hspace{0.5cm} $M_{l+1} \gets \textbf{MaskGen}(r, T_r, P_{l+1})$
\STATE \hspace{0.5cm} // Detect rank changes per round:
\STATE \hspace{0.5cm} $N_t \gets \textbf{RankDet}(M_{l+1})$
\STATE \hspace{0.5cm} // Prune uploading parameters:
\STATE \hspace{0.5cm} $\hat{P}_{l+1} \gets \textbf{CommPru}(P_{l+1}, M_l)$
\STATE \hspace{0.5cm} \textbf{return} $\hat{P}_{l+1}, M_{l+1}$
\end{algorithmic}
\end{algorithm}

Algorithm \ref{alg1} outlines the process of federated adaptive rank allocation. It starts with parameter initialization and client selection (\textit{line 4-6}). The server prunes parameters using global rank masks via CommPru (\textit{lines 7-8}, see Section~\ref{commpru}) and broadcasts them to the clients (\textit{line 9}). Clients perform parallel local training (\textit{lines 10-11}), updating parameters based on local datasets (\textit{lines 21-22}). Local rank masks are generated from updated parameters using MaskGen (\textit{lines 23-24}, see Section~\ref{maskgen}), determining the number of trainable parameters via RankDet (\textit{lines 25-26}, see Section~\ref{method3}). During uploading, global rank masks guide parameter pruning via CommPru (\textit{lines 27-28}, see Section~\ref{commpru}). After receiving client updates, the server aggregates parameters using the FedAvg algorithm \cite{mcmahan2017communication} (\textit{lines 12-13}), refines global rank masks using threshold arbitration via FedArb (\textit{lines 14-15}, see Section~\ref{fedarb}), and generates global parameters. Finally, model accuracy is assessed using a test dataset (\textit{lines 17-19}).

\section{Design Details of FedARA}
\label{design}

In this section, we present the design of FedARA, consisting of three interrelated methods: truncated SVD adaptation, dynamic rank allocation, and rank-based module pruning.

\subsection{Truncated SVD Adaptation}\label{method1}
In previous work, LoRA \cite{hu2022lora} decomposes the trainable weight matrix into pre-trained weights and incremental weights as:
\begin{equation}
W = W_{\text{pre}} + \Delta W = W_{\text{pre}} + \frac{\alpha}{r} BA,
\end{equation}
where $W_{\text{pre}}, \Delta W \in \mathbb{R}^{d_{\text{out}} \times d_{\text{in}}}$, $B \in \mathbb{R}^{d_{\text{out}} \times r}$, $A \in \mathbb{R}^{r \times d_{\text{in}}}$, and $\frac{\alpha}{r}$ is a scaling factor with $r \ll \min\{d_{\text{out}}, d_{\text{in}}\}$.
During training, only the low-rank matrices $A$ and $B$ are updated, while $W_{\text{pre}}$ remains frozen. This decomposition improves parameter efficiency when $r(d_{\text{out}} + d_{\text{in}}) < d_{\text{out}} d_{\text{in}}$. Typically, $A$ is initialized with Gaussian noise, and $B$ is initialized to zero, ensuring $\Delta W = 0$ at initialization.

However, our pre-experiments and prior studies \cite{babakniya2023slora, yan2024federa} reveal a significant performance gap in FedLoRA between IID and non-IID settings. We hypothesize that this gap stems from the inherent structure and asymmetric initialization of LoRA.
Specifically, LoRA's low-rank structure offers a high degree of flexibility but inherently couples magnitude and directional updates within the same matrix, amplifying client divergence under data heterogeneity. Additionally, its asymmetric initialization introduces randomness and bias in early training, further intensifying client drift in non-IID settings.

To address this issue, we draw inspiration from the structure of truncated Singular Value Decomposition (SVD), which decomposes a matrix into the singular values and vectors of its low-rank subspace. Accordingly, we propose a truncated SVD adaptation method that mirrors this structure, formulated as
\begin{equation}
    \label{eq:tsvd}
    W = W_{\text{pre}} + \Delta W \;=\; W_{\text{pre}} \;+\; \frac{\alpha}{r} B E A,
\end{equation}
where \(E \in \mathbb{R}^{r \times r}\) is a diagonal matrix, and \(B \in \mathbb{R}^{d_{\text{out}} \times r}\), \(A \in \mathbb{R}^{r \times d_{\text{in}}}\). Both \(A\) and \(B\) are initialized with random Gaussian values, while \(E\) is initialized to zero, ensuring that \(\Delta W = 0\) at the beginning of training. In contrast to the standard truncated SVD, we do not impose orthogonality constraints on \(B\) and \(A\), as such constraints reduce flexibility and incur additional computational overhead in each training round.

\begin{figure}[!t]
  \centering
  \includegraphics[width=0.9\linewidth]{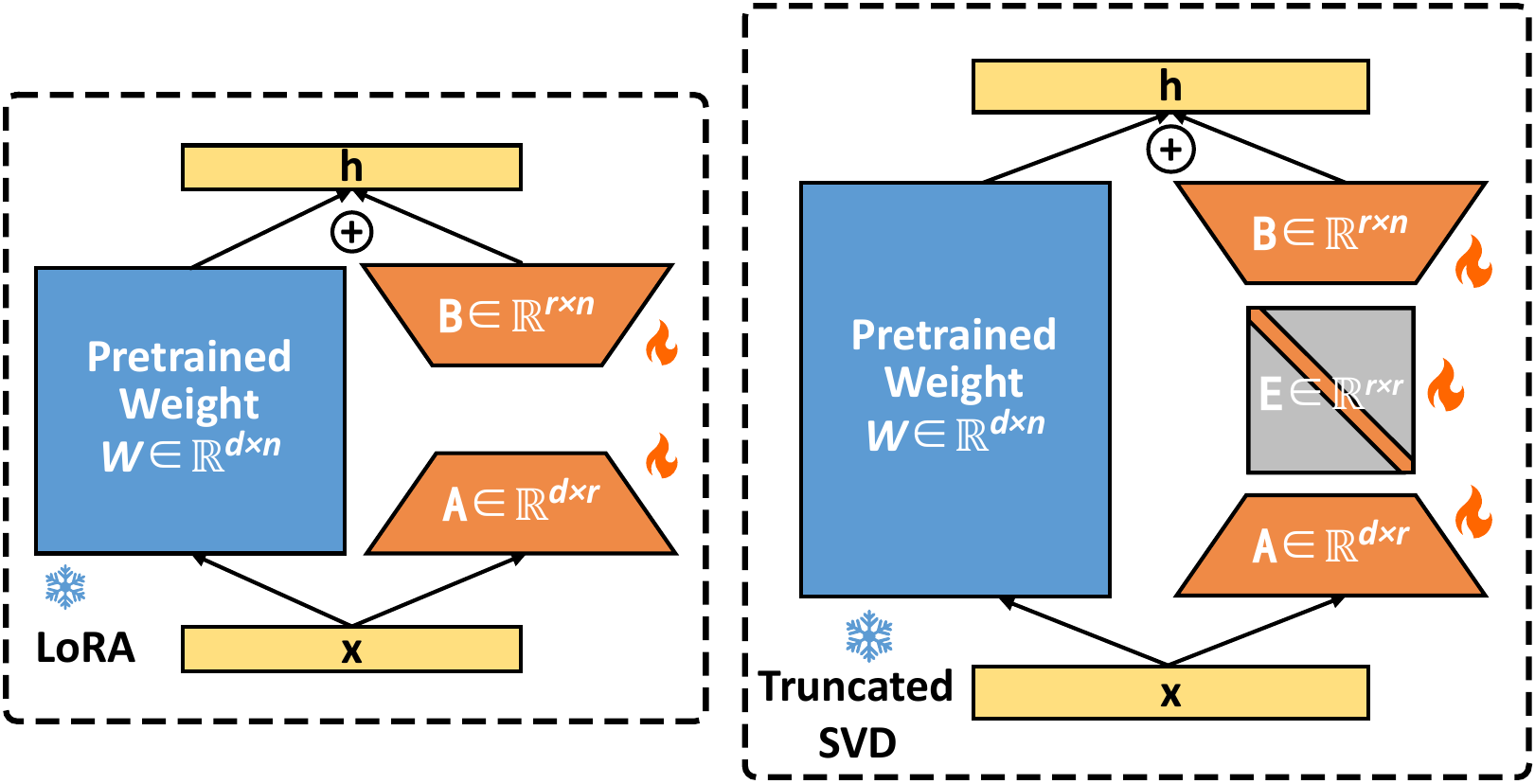}
  \caption{Illustration of truncated SVD adaptation, where a diagonal matrix \(E\) is inserted between the LoRA matrices.}
  \label{fig4}
\end{figure}
As illustrated in Figure~\ref{fig4}, a diagonal matrix \(E\) is inserted between the low-rank subspaces at negligible cost (\(r \ll \min(d_{\text{in}}, d_{\text{out}})\). The diagonal structure of \(E\) restricts updates to independent scaling along each latent dimension, which helps mitigate inter-client variability under non-IID data. At the same time, symmetric initialization ensures balanced gradient flow, offering clients a stable starting point and enhancing global consistency during federated fine-tuning.

Next, we provide a simplified theoretical analysis to offer intuition for the effectiveness of truncated SVD adaptation. For each client $k$, the rank-$r$ update is defined as:
\begin{equation}
\Delta W_k^{\text{BA}}
   =\sum_{i=1}^{r} b_{k,i} a_{k,i}^{\top}, 
\quad
\Delta W_k^{\text{BEA}}
   =\sum_{i=1}^{r} e_{k,i} b_{k,i} a_{k,i}^{\top},
\end{equation}
and we quantify client heterogeneity by the drift variance, assuming zero-mean updates:
\begin{equation}
\sigma^{2}
=\frac{1}{K}\sum_{k=1}^{K}\!
   \bigl\Vert \Delta W_k-\overline{\Delta W}\bigr\Vert_{F}^{2}
\;\approx\;
\mathbb{E}\!\left[\!\Vert\Delta W\Vert_{F}^{2}\right].
\end{equation}

Let $T_i$ denote the $i$-th rank-1 term. Then, the expression $\Vert\Delta W\Vert_{F}^{2}$ can be reformulated as:
\begin{equation}
\Vert\Delta W\Vert_{F}^{2}
   = \sum_{i=1}^{r} \sum_{j=1}^{r} \langle T_i, T_j \rangle,
\end{equation}
where \(\langle T_i, T_j \rangle = \operatorname{Tr}(T_i^{\top} T_j)\). Using the identity \(\langle uv^{\top}, xy^{\top} \rangle = (u^{\top}x)(v^{\top}y)\), the inner product becomes:
\begin{equation}
\langle T_i, T_j \rangle =
\begin{cases}
(b_i^{\top} b_j)(a_i^{\top} a_j), & \text{(BA)} \\
e_i e_j\,(b_i^{\top} b_j)(a_i^{\top} a_j), & \text{(BEA)}
\end{cases}
\quad,
\label{eq:inner}
\end{equation}
\rev{To make the analysis tractable, we adopt a simplified separable-covariance assumption:}
\begin{equation}
\mathbb{E}[b_i^{\top} b_j]=
\begin{cases}
\tau_b, & i = j \\
\rho_b, & i \ne j
\end{cases} 
\quad,
\quad
\mathbb{E}[a_i^{\top} a_j]=
\begin{cases}
\tau_a, & i = j \\
\rho_a, & i \ne j
\end{cases}
\quad,
\end{equation}
\rev{where $\rho_a\rho_b\neq0$, capturing non-trivial cross-rank covariances.
This simplified model is inspired by separable (Kronecker-factored) covariance approximations~\cite{martens2015optimizing}.}
For the truncated SVD adaptation, we further assume:
\begin{equation}
\mathbb{E}[e_i] = 0, \quad
\mathbb{E}[e_i^2] = \tau_e, \quad
\bigl|\mathbb{E}[e_i e_j]\bigr| \le \frac{c}{r} \quad (i \ne j).
\label{eq:gate}
\end{equation}

Each $e_i$ is initialized as $e_i^{(0)} = 0$ and updated by $e_i \leftarrow e_i - \eta \langle \nabla \ell(x; W),\, b_i a_i^{\top} \rangle$, where $e_i$ only interacts with its own rank-1 component, leading to independent gradient noise across $i \ne j$. Under symmetric mini-batch sampling, the noise is unbiased with $\mathbb{E}[\Delta e_i] = 0$. Additionally, with Gaussian initialization, the inner product $(b_i^{\top} b_j)(a_i^{\top} a_j) = O(1/\sqrt{d})$, yielding residual statistical dependence of order $O(1/r)$ when $r<\sqrt d$.

Substituting equations~\eqref{eq:inner}-\eqref{eq:gate}, we obtain for LoRA:
\begin{equation}
\label{eq:lora}
\mathbb{E}\!\left[\|\Delta W^{\text{BA}}\|_F^2\right] = r\tau_b\tau_a + r(r-1)\rho_b\rho_a = \Theta(r^2), 
\end{equation}
while for truncated SVD adaptation, the $r(r - 1)$ cross terms are scaled by $\mathbb{E}[e_i e_j]$ and vanish up to order $O(r)$, yielding:
\begin{equation}
\label{eq:svd}
\mathbb{E}\!\left[\|\Delta W^{\text{BEA}}\|_F^2\right] = r\tau_e\tau_b\tau_a + O(r) = \Theta(r).   
\end{equation}
\rev{Thus, the variance scales as \(\sigma^2_{\text{BA}} = \Theta(r^2)\) and \(\sigma^2_{\text{BEA}} = \Theta(r)\), indicating that the diagonal matrix $E$ suppresses quadratic cross-rank covariances induced by non-IID updates. This result is derived under the simplified separable-covariance model. When cross-rank covariances vanish ($\rho_a\rho_b = 0$), all cross terms disappear and both methods yield \(\Theta(r)\) variance, leaving little room for further improvement. Accordingly, our analysis focuses on the regime with non-zero cross-rank covariances ($\rho_a\rho_b \neq 0$) to examine the effect of truncated SVD adaptation, while acknowledging that this simplified model may not strictly hold in practice.}

\begin{figure}[!t]
  \centering
  \includegraphics[width=\linewidth]{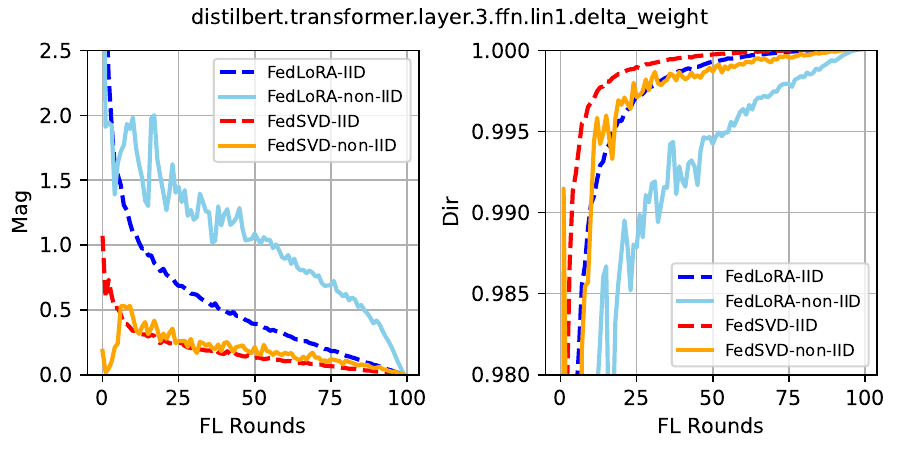}
  \caption{Magnitude (left) and direction (right) discrepancies between global and local models under different FL structures. FedLoRA and FedSVD correspond to fine-tuned models utilizing LoRA and truncated SVD adaptation, respectively, with a rank of 12.}
  \label{fig5}
\end{figure}

To further understand the difference in global weight updates between LoRA and truncated SVD adaptation in FL, we introduce \textbf{magnitude} and \textbf{directional} discrepancies between the global model $\boldsymbol{\theta}_{\text{global}}$ and the local models $\boldsymbol{\theta}_{\text{local}}^{(i)}$ across selected clients. The magnitude discrepancy is defined as:
\begin{equation}
\text{Mag} = \sum_{i=1}^{K} \left\| \boldsymbol{\theta}_{\text{global}} - \boldsymbol{\theta}_{\text{local}}^{(i)} \right\|_F,
\end{equation}
where $K$ denotes the number of selected clients per round, and $\|\cdot\|_F$ represents the Frobenius norm. This metric captures the magnitude difference between the global and local models. The directional discrepancy is given by:
\begin{equation}
\text{Dir} = \frac{1}{{K}} \sum_{i=1}^{{K}} \frac{\langle \boldsymbol{\theta}_{\text{global}}, \boldsymbol{\theta}_{\text{local}}^{(i)} \rangle}{\|\boldsymbol{\theta}_{\text{global}}\|_F \, \|\boldsymbol{\theta}_{\text{local}}^{(i)}\|_F}.
\end{equation}
This metric measures the average cosine similarity between the global and local models, with values closer to 1 indicating higher directional alignment. 

We fine-tune the \textit{DistilBERT} \cite{sanh2019distilbert} model on the \textit{20News} \cite{lang1995newsweeder} dataset in an FL setting with 100 rounds, selecting 10 clients per round. As a representative, we extract the delta weights of \texttt{layer.3.ffn.lin1}, which reconstructs the low-rank module into an incremental matrix matching the dimensions of the pre-trained model.
As shown in Figure~\ref{fig5}, FedSVD shows greater consistency in global model updates compared to FedLoRA under non-IID settings. FedLoRA exhibits significant sensitivity to heterogeneity, manifesting in pronounced magnitude and direction discrepancies, while FedSVD demonstrates much smaller fluctuations. This highlights that the truncated SVD structure is more effective than LoRA in representing similar features across diverse clients.

\subsection{Dynamic Rank Allocation}\label{method2}
The fixed-rank configuration assigns uniform ranks to modules across different layers and components, resulting in communication inefficiency due to its inability to adapt to the varying contributions of layers and components to model performance. This inefficiency manifests in two aspects: the fixed rank allocation yields suboptimal performance under similar communication overhead and requires higher communication costs to achieve comparable performance with the dynamic rank allocation.

Inspired by AdaLoRA \cite{zhang2023adalora}, which introduces a framework for centralized rank assignment, we propose a dynamic rank allocation method in FL that efficiently utilizes limited parameters to prioritize the most critical weights, thereby enhancing communication efficiency. This method progressively identifies the most significant weights while pruning less important parameters across clients, involving three key steps: \textit{Generation of Local Rank Masks}, \textit{Arbitration of Global Rank Masks}, and \textit{Communication with Arbitrated Masks}.

\begin{figure}[!t]
  \centering
  \includegraphics[width=\linewidth]{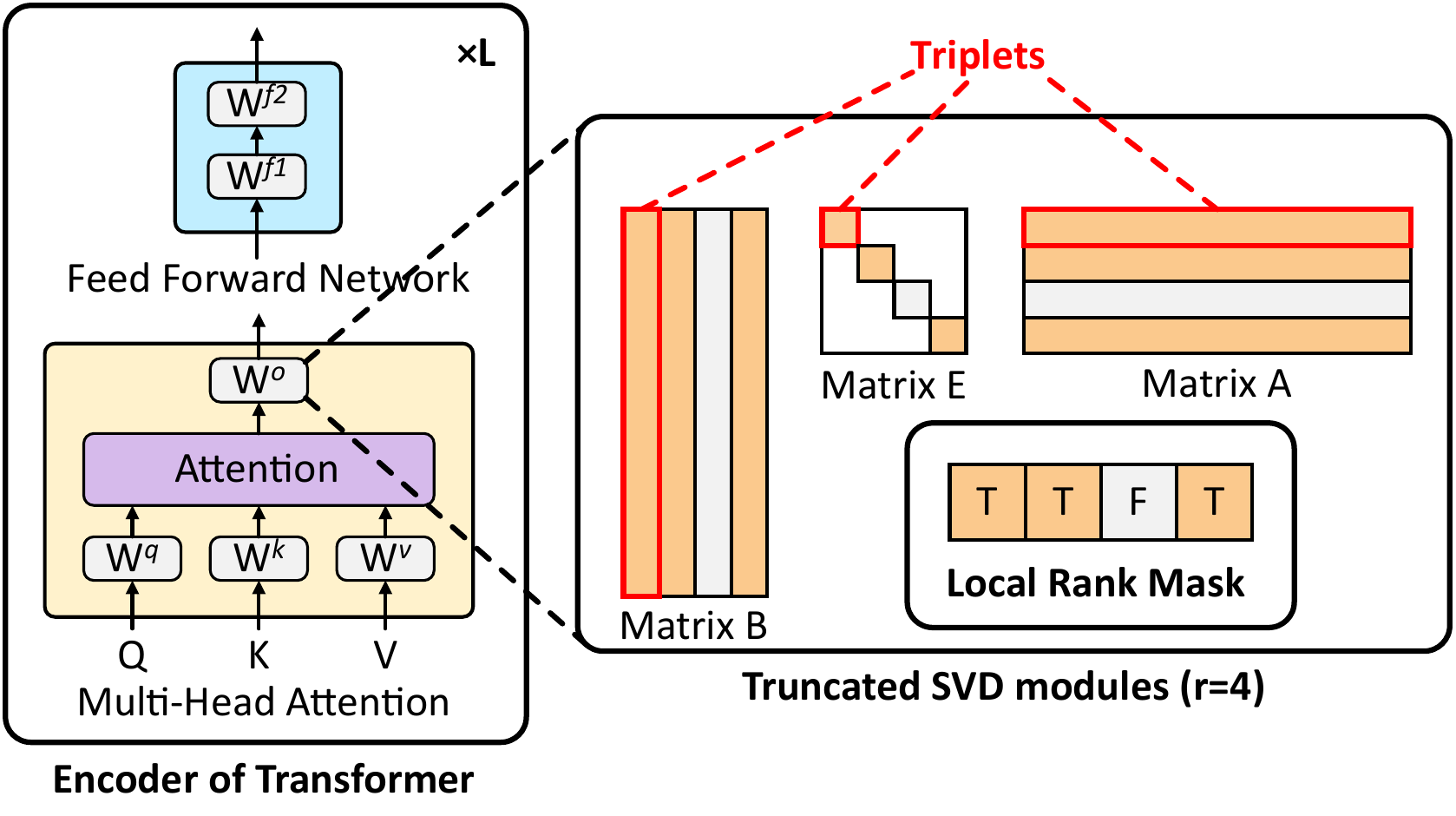}
  \caption{Construction of triplets in SVD modules and the structure of local rank masks.}
  \label{fig6}
\end{figure}

\subsubsection{Generation of Local Rank Masks (MaskGen)}\label{maskgen}
During federated fine-tuning, local rank mask generation is directly influenced by the rank budget, defined as the total sum of ranks across all layers and components. This budget is progressively reduced, eventually approaching the pre-set average target rank. The relationship between the budget \(b^{(t)}\) and the FL round \(t\) is given by:
\begin{equation}
b^{(t)} =
\begin{cases}
b^{(0)}, & 0 \leq t < t_w, \\
b^{(T)} + b^{(c)} \left( 1 - \frac{t - t_w - t_f}{T - t_w - t_f} \right)^3, & t_w \leq t < T - t_f, \\
b^{(T)}, & \text{otherwise}.
\end{cases}
\end{equation}
where \(b^{(t)} = b^{(0)} - b^{(R)}\), \(t_w\) represents the warm-up rounds, \(t_f\) indicates the final stabilized rounds, \(T\) denotes the total rounds.

Local clients identify relatively important rank candidates based on the reduced rank budget. The importance of each rank is evaluated using triplets, as shown in Figure~\ref{fig6}, which consist of the diagonal value in \(E\), the left vector in \(B\), and its corresponding right vector transpose in \(A\). This can be represented as:
\begin{equation}
I_{n,i} = I(E_{n,i}) + \frac{1}{d_1} \sum_{j=1}^{d_1} I(B_{n,ji}) + \frac{1}{d_2} \sum_{j=1}^{d_2} I(A_{n,ij}),
\end{equation}
where \(I(E_{n,i})\), \(I(B_{n,ji})\), and \(I(A_{n,ij})\) denote the importance scores of the corresponding parts. As shown in Table~\ref{tab:ipt}, we compare four importance score evaluation strategies:
\textbf{Mag} denotes the weight magnitude, \(I(w_{i,j}) = |w_{i,j}|\); 
\textbf{Grad} represents the gradient, \(|\text{grad}(w_{i,j})|\); 
\textbf{Mixed} combines both weight and gradient information via their product, \(|\text{grad}(w_{i,j}) \cdot w_{i,j}|\); 
and \textbf{Sensitivity} corresponds to the sensitivity-based method proposed in AdaLoRA~\cite{zhang2023adalora}. 
Among the evaluated strategies, the magnitude-based method consistently achieves the best performance across different datasets and models, outperforming both the mixed strategy and gradient-based approaches. While the sensitivity-based method also improves accuracy, it introduces approximately 30\% additional computational overhead. To avoid the high cost of sensitivity and gradient-based estimation, we adopt the magnitude-based score as a simple yet effective alternative.

\begin{table}[t]
\centering
\caption{Performance and local computation overhead under different importance score evaluation methods (non-IID alpha=0.1). Results are reported in percentage (\%).}
\label{tab:ipt}
\begin{tabular}{lccccc}
\toprule
\multirow{2}{*}{\textbf{Methods}} & \multicolumn{2}{c}{\textbf{DistilBERT}} & \multicolumn{2}{c}{\textbf{BERT}} & \multirow{2}{*}{\textbf{Comp.}} \\
  & \textbf{20News} & \textbf{Semeval} & \textbf{20News} & \textbf{Semeval} & \\
\midrule
\textbf{Mag}         & \textbf{63.89} & \textbf{80.24} & \textbf{64.32} & \textbf{76.89} & 1$\times$ \\
\textbf{Grad}        & 63.31 & 79.59 & 63.68 & 76.05 & 0.97$\times$ \\
\textbf{Mixed}   & 63.79 & 79.87 & 63.94 & 76.42 & 1.11$\times$ \\
\textbf{Sensitivity} & \textbf{63.89} & 80.06 & 64.21 & 76.70 & 1.30$\times$ \\
\bottomrule
\end{tabular}
\end{table}

As illustrated in Figure~\ref{fig6}, each SVD module initially contains \(r\) triplets, which can be inserted into six different components. The local rank mask of each client is generated based on the sorting of all triplets. Specifically, if the importance of a rank falls within the \(\textit{top-}b^{(t)}\), it is marked as \texttt{True} in the mask; otherwise, it is marked as \texttt{False}. Each client uploads these masks to the server along with the parameters.

\subsubsection{Arbitration of Global Rank Masks (FedArb)}\label{fedarb}
After collecting all candidates of local rank masks from all selected clients, the server determines global rank masks by arbitrating the fraction of clients reporting \texttt{True} at each position against a preset threshold $T_h$. This arbitration process is given by:
\begin{equation}
\label{eq:global-mask}
M_{\mathrm{global}}^{(i)} 
=
\begin{cases}
True, & \text{if } 
\displaystyle 
\frac{1}{\lvert{K}\rvert}
\sum_{k \in {K}} M_{k}^{(i)} 
> T_h,\\
False, & \text{otherwise}.
\end{cases}
\end{equation}
here, ${K}$ denotes the set of participating clients. The arbitrated global rank masks are broadcast to all clients and used to optimize communication parameters between the server and clients. 
As shown in Table~\ref{tab:arb}, we compare two arbitration strategies: \textbf{FedARA} collects local rank masks from clients and performs arbitration on the server; and \textbf{FedARA-global} generates global rank masks directly from the aggregated model. Although the communication overhead is similar for both strategies, FedARA achieves better performance on the 20News dataset~\cite{lang1995newsweeder}, highlighting the effectiveness of leveraging local rank information for centralized arbitration.

\begin{table}[t]
\centering
\caption{ Performance and communication overhead under different arbitration strategies (non-IID alpha=0.1). Results are reported in percentage(\%).}
\label{tab:arb}
\begin{tabular}{lccccc}
\toprule
\multirow{2}{*}{\textbf{Methods}} & \multicolumn{2}{c}{\textbf{20News}} & \multirow{2}{*}{\textbf{Comm.}} \\
  & \textbf{DistilBERT} & \textbf{BERT} & \\
\midrule
\textbf{FedARA}               & \textbf{62.94} & \textbf{63.52} & 8.80 GB \\
\textbf{FedARA-global}        & 62.78 & 63.36 & 8.82 GB \\
\bottomrule
\end{tabular}
\end{table}

\subsubsection{Communication with Arbitrated Masks (CommPru)}\label{commpru}
The global and local rank masks are transmitted alongside the parameters, yet their communication overhead remains negligible due to their boolean nature and significantly smaller size relative to the full parameters. During server-client communication, if a global rank mask in the truncated SVD modules is marked as \texttt{False}, the corresponding parameters are pruned. The complete weight matrix is reconstructed on the client or server using global rank masks. This approach substantially reduces communication overhead in each round.

FedARA progressively achieves adaptive rank allocation during federated fine-tuning rounds by following the aforementioned steps, ultimately preserving only the most important ranks while removing less significant ones. This process significantly improves communication efficiency.

\subsection{Rank-based Module Pruning}\label{method3}
Although the dynamic parameter allocation method with truncated SVD modules reduces communication overhead in FL by limiting the upload and download of less critical ranks, it does not decrease the number of trainable parameters during local training. This limitation arises because frameworks such as PyTorch and Transformers manage trainable parameters at the matrix level, preventing selective updates at the rank level. Fine-grained sparse gradient updates require additional computation overhead. To address this, we introduce a rank-based module pruning mechanism (RankDet in Algorithm~\ref{alg1}) that safely removes inactive SVD modules in a structured manner.

Specifically, each SVD module on the local client has a rank parameter that changes over time according to local training and the ranking results of triplets. Dynamic rank allocation ensures that the ranks of SVD modules either remain constant or gradually decrease as FL training progresses. Critical ranks are preserved, while less significant ones are progressively pruned. Once a rank is reduced to a predefined threshold (set to zero in this work), it is deemed non-essential and can be safely removed.

Thus, FedARA continuously monitors the rank changes of each inserted SVD module in every FL round. When a module's rank falls below the threshold, all three matrices in the SVD module become non-trainable. This process does not affect model performance, as these inactive ranks are excluded from the communication and aggregation. By eliminating unnecessary training of specific SVD modules, this method reduces computational overhead and memory consumption, making it particularly advantageous for deployment on resource-constrained devices.

\section{Experimental Setups}
\label{setup}
FedARA extends AdaLoRA \cite{zhang2023adalora}, originally designed for dynamic parameter configuration in centralized training, to the federated setting. Built on Hugging Face's Transformers library, it supports a broad range of pre-trained NLP models. FedARA adopts a random client selection mechanism, which can be substituted with more advanced strategies \cite{zou2024value, fu2023client}. Communication between clients and the server is synchronized via weighted averaging \cite{mcmahan2017communication}, maintaining compatibility with diverse optimization and aggregation methods \cite{li2020federated, karimireddy2020scaffold, mills2023fedgbo, yao2024fedgkd, qi2023model}. Moreover, FedARA incorporates model quantization techniques such as QLoRA \cite{dettmers2023qlora} and QA-LoRA \cite{xu2024qalora} to reduce computational and memory overhead, potentially lowering communication costs.

\begin{table}[t]
\begin{center}
  \caption{Different dataset settings and sizes in sequence classification and summarization tasks of NLP.}
  \label{tab:datasets}
  \begin{tabular}{ccccc}
    \toprule
     \textbf{Tasks} & \textbf{Datasets}&\textbf{\makecell{Client \\ Num.}}&\textbf{Labels}&\textbf{Samples}\\
    \midrule
    \multirow{4}{*}{\makecell{Classification}} 
    & 20News \cite{lang1995newsweeder} & 100 & 20 & 18.8k\\
    & Semeval \cite{hendrickx2010semeval} & 100 & 19 & 10.7k \\
    & AG News \cite{zhang2015character} & 1000 & 4 & 127.6k\\
    & News Category \cite{misra2022news} & 1000 & 15 & 148.1k \\
    \midrule
    \multirow{1}{*}{\makecell{Summarization}} 
    & CNN/DailyMail \cite{hermann2015teaching} & 1000 & - & 287k \\
    \bottomrule
  \end{tabular}
\end{center}  
\end{table}

\textbf{Metrics.} The primary evaluation metrics are final accuracy for classification tasks, ROUGE scores~\cite{lin2004rouge} for summarization tasks, and total communication overhead after a fixed number of FL rounds, ensuring fair performance comparison. Secondary metrics include per-round communication cost, adaptive rank allocation, total training time, memory footprint, and energy consumption on edge devices. These metrics offer a comprehensive evaluation of system efficiency and effectiveness.

\textbf{Hardware.} Following prior FL research~\cite{cai2023efficient}, all experiments are conducted via emulation on a laptop equipped with an RTX 4070 GPU. Local training time is measured on three edge devices, AGX Orin, Orin Nano, and Raspberry Pi 5, chosen for their representativeness of common resource-constrained hardware. These results are integrated into our emulation framework to estimate total training time in a homogeneous FL setting. In line with previous studies~\cite{cai2023efficient}, we assume a communication bandwidth of 1MB/s between the server and clients, reflecting realistic conditions across IoT, home WiFi, and cellular networks.

\textbf{Datasets and Partitioning.} FedARA is evaluated on sequence classification and summarization tasks using five diverse NLP datasets, as summarized in Table~\ref{tab:datasets}. 
\textit{Classification tasks} include:  
(1) \textit{20News}~\cite{lang1995newsweeder}, comprising documents from 20 different newsgroups;  
(2) \textit{Semeval}~\cite{hendrickx2010semeval}, covering various NLP and semantic evaluation tasks;  
(3) \textit{AG News}~\cite{zhang2015character}, containing news articles from four categories;  
(4) \textit{News Category}~\cite{misra2022news}, a dataset of diverse news headlines, where we focus on the top 15 most frequent labels, covering approximately 70\% of the data.  
\textit{Summarization task:}  
(5) \textit{CNN/DailyMail}~\cite{hermann2015teaching}, a widely used dataset for abstractive summarization.
Each classification dataset is split into training, validation, and test sets in an 8:1:1 ratio. We fine-tune a \textit{DistilBERT} model on the \textit{News Category} dataset (top-15 labels) and use the resulting model to generate pseudo-labels for the summarization dataset.
To simulate non-IID conditions in a federated setup, we adopt Dirichlet-based partitioning with $\alpha \in \{1, 0.1, 0.01\}$. Following FedAvg~\cite{mcmahan2017communication}, we also consider a pathological non-IID setting where each client holds samples from only 1-2 labels. For IID settings, we set $\alpha = 1000$ to approximate a uniform label distribution across clients.

\textbf{Model.} We adopt three widely used transformer-based models for NLP tasks. For classification, we use \textit{BERT-base}~\cite{devlin2019bert}, which consists of 12 transformer layers, and its compressed variant \textit{DistilBERT-base}~\cite{sanh2019distilbert}, which retains most of \textit{BERT}'s performance with only 6 layers via knowledge distillation. For summarization, we employ \textit{BART-base}~\cite{lewis2019bart}, an encoder-decoder model designed for sequence-to-sequence generation. All selected models are based on the transformer architecture, which remains foundational for even the latest NLP models~\cite{dubey2024llama}. Our work focuses on federated adaptive rank allocation for transformer models. While this study evaluates \textit{BERT}, \textit{DistilBERT}, and \textit{BART} across classification and summarization tasks to ensure generality and feasibility on resource-constrained devices, our method is readily applicable to newer architectures and broader NLP tasks.

\begin{table*}[htbp]
  \caption{Comparison of FedPEFT methods across different models and datasets. In N1 (\textcolor{black}{N2}), N1 represents the accuracy under non-IID data (Pathological distribution), while \textcolor{black}{N2} denotes the relative accuracy degradation compared to IID data.}
  \label{tab:main}
  \centering
  \begin{tabular}{cccllll}
    \toprule
    \textbf{Models} & \textbf{Methods} & \textbf{\makecell{PEFT Comm \\ Overhead (GB)}} & \textbf{\makecell{20News (\%)}} & \textbf{\makecell{Semeval (\%)}} & \textbf{\makecell{AG \\ News (\%)}} & \textbf{\makecell{News \\ Category (\%)}} \\
    \midrule
    \multirow{9}{*}{DistilBERT} 
    & FedLoRA (r=6) & 11.12 & 50.81 (\textcolor{black}{↓14.19}) & 34.76 (\textcolor{black}{↓45.71}) & 86.74 (\textcolor{black}{↓1.64}) & 56.34 (\textcolor{black}{↓14.67}) \\
    & FedAdapter-h (a=27) & 11.34 & 51.97 (\textcolor{black}{↓13.06}) & 19.09 (\textcolor{black}{↓62.57}) & 85.93 (\textcolor{black}{↓2.60}) & 59.29 (\textcolor{black}{↓12.03}) \\
    & FedAdapter-p (a=54) & 11.24 & 52.19 (\textcolor{black}{↓12.76}) & 20.04 (\textcolor{black}{↓60.93}) & 86.33 (\textcolor{black}{↓2.56}) & 59.47 (\textcolor{black}{↓11.99}) \\
    & SLoRA (r=6) & 11.12 & 48.83 (\textcolor{black}{↓15.33}) & 31.74 (\textcolor{black}{↓47.01}) & 85.68 (\textcolor{black}{↓2.17}) & 55.57 (\textcolor{black}{↓14.91}) \\
    & FeDeRA (r=6) & 11.12 & 52.28 (\textcolor{black}{↓11.43}) & 28.33 (\textcolor{black}{↓50.95}) & 86.22 (\textcolor{black}{↓1.94}) & 56.45 (\textcolor{black}{↓14.64}) \\
    & FFA-LoRA-dr (r=12) & 11.12 & 53.96 (\textcolor{black}{↓10.99}) & 31.91 (\textcolor{black}{↓48.46}) & \underline{87.34} (\textcolor{black}{↓1.55}) & 61.58 (\textcolor{black}{↓10.14}) \\
    & FFA-LoRA (r=6) & \underline{5.56} & 53.52 (\textcolor{black}{↓11.24}) & 29.80 (\textcolor{black}{↓50.29}) & 87.00 (\textcolor{black}{↓1.70}) & 61.27 (\textcolor{black}{↓10.36}) \\
    & \textbf{FedARA (init r=12)} & 9.29 & \textbf{55.05} (\textbf{\textcolor{black}{↓10.03}}) & \textbf{56.68} (\textbf{\textcolor{black}{↓24.29}}) & \textbf{87.40} (\textbf{\textcolor{black}{↓1.37}}) & \underline{63.47} (\underline{\textcolor{black}{↓8.10}}) \\
    & \textbf{FedARA (init r=6)} & \textbf{4.64} & \underline{54.86} (\underline{\textcolor{black}{↓10.22}}) & \underline{55.77} (\underline{\textcolor{black}{↓25.20}}) & 87.27 (\underline{\textcolor{black}{↓1.50}}) & \textbf{63.56} (\textbf{\textcolor{black}{↓8.01}}) \\
    \midrule
    \multirow{9}{*}{BERT} 
    & FedLoRA (r=6) & 22.24 & 46.27 (\textcolor{black}{↓18.03}) & 34.91 (\textcolor{black}{↓45.35}) & 86.47 (\textcolor{black}{↓2.44}) & 55.70 (\textcolor{black}{↓16.09}) \\
    & FedAdapter-h (a=27) & 22.68 & 49.55 (\textcolor{black}{↓15.31}) & 39.20 (\textcolor{black}{↓41.15}) & 86.40 (\textcolor{black}{↓2.52}) & 56.34 (\textcolor{black}{↓15.50}) \\
    & FedAdapter-p (a=54) & 22.46 & 49.36 (\textcolor{black}{↓16.04}) & 31.07 (\textcolor{black}{↓49.14}) & 85.66 (\textcolor{black}{↓3.41}) & 57.68 (\textcolor{black}{↓15.14}) \\
    & SLoRA (r=6) & 22.24 & 45.60 (\textcolor{black}{↓16.73}) & 28.42 (\textcolor{black}{↓47.93}) & 84.36 (\textcolor{black}{↓3.37}) & 54.20 (\textcolor{black}{↓15.76}) \\
    & FeDeRA (r=6) & 22.24 & 50.58 (\textcolor{black}{↓13.95}) & 31.69 (\textcolor{black}{↓49.02}) & 86.97 (\textcolor{black}{↓2.42}) & 54.82 (\textcolor{black}{↓17.72}) \\
    & FFA-LoRA-dr (r=12) & 22.24 & 54.77 (\textbf{\textcolor{black}{↓9.15}}) & 46.51 (\textcolor{black}{↓34.16}) & \underline{87.39} (\textcolor{black}{↓1.68}) & 65.27 (\textcolor{black}{↓6.91}) \\
    & FFA-LoRA (r=6) & \underline{11.12} & 53.10 (\textcolor{black}{↓11.33}) & 38.25 (\textcolor{black}{↓42.23}) & 86.67 (\textcolor{black}{↓1.45}) & 62.60 (\textcolor{black}{↓9.81}) \\
    & \textbf{FedARA (init r=12)} & 18.86 & \textbf{55.61} (\underline{\textcolor{black}{↓9.45}}) & \textbf{48.67} (\textbf{\textcolor{black}{↓31.03}}) & \textbf{87.40} (\textbf{\textcolor{black}{↓1.27}}) & \underline{65.48} (\underline{\textcolor{black}{↓6.41}}) \\
    & \textbf{FedARA (init r=6)} & \textbf{9.46} & \underline{55.08} (\textcolor{black}{↓9.98}) & \underline{48.35} (\underline{\textcolor{black}{↓31.35}}) & 87.32 (\underline{\textcolor{black}{↓1.35}}) & \textbf{65.71} (\textbf{\textcolor{black}{↓6.18}}) \\
    \bottomrule
  \end{tabular}
\end{table*}

\textbf{Baselines.} We compare FedARA against several recent FedPEFT methods:
\begin{itemize}
    \item \textbf{FedLoRA}: a widely adopted baseline with extensive variants and empirical studies~\cite{zhang2024towards, wang2024flora, yang2024sa, bai2024federated}.
    \item \textbf{FedAdapter-h}: incorporates Adapter-h~\cite{houlsby2019parameter} modules into the attention and feed-forward layers of transformers.
    \item \textbf{FedAdapter-p}: inserts adapter modules only in the feed-forward layers~\cite{pfeiffer2021adapterfusion}. We exclude AdaFL~\cite{cai2023efficient} as it is a parameter-efficient variant of FedAdapter-p and does not explicitly address non-IID data heterogeneity.

    \item \textbf{SLoRA}: employs a two-stage training scheme~\cite{babakniya2023slora}, where the first stage applies sparse fine-tuning to improve initialization for LoRA in the second stage. We allocate 10\% of the total FL rounds to the sparse stage.
    \item \textbf{FeDeRA}: initializes LoRA using singular value decomposition of pre-trained weights~\cite{yan2024federa}.
    \item \textbf{FFA-LoRA}: updates only the \(B\) matrix in LoRA while freezing \(A\)~\cite{zhang2023lora, sun2024improving, zhu2024asymmetry}. To mitigate reduced expressiveness, we also evaluate \textbf{FFA-LoRA-dr}, which employs orthogonal initialization for \(A\) and doubles the rank.
\end{itemize}

\textbf{Hyperparameters.} Learning rates are selected via grid search in the range of 1e-5 to 5e-3, depending on the dataset and model. All methods use a batch size of 4, one epoch per FL round, a maximum sequence length of 64, and are fine-tuned for 100 rounds with 10 clients per round (10\% or 1\% selection). 
For summarization, we set the source length to 512, the target length to 64, and the beam size to 1 to fit edge device constraints. Adaptive rank allocation starts after 5 warm-up rounds and decays until round 50, targeting one-quarter of the original rank with a global mask threshold of 0.5.
Learning rates decay linearly across rounds, using the Adam optimizer. The LoRA scaling factor $\alpha$ is fixed at 16, following prior work~\cite{hu2022lora, zhang2023adalora}.

\section{Evaluation}
\label{evaluation}
\subsection{Performance under non-IID data}
We conduct experiments on classification tasks under pathological non-IID and IID settings, using two different models across four datasets. The results are summarized in Table~\ref{tab:main}, where the best performance is highlighted in \textbf{bold} and the second-best is \underline{underlined}. Each result is averaged over five runs with different random seeds. 

\begin{figure}[t]
\centering
\includegraphics[width=\linewidth]{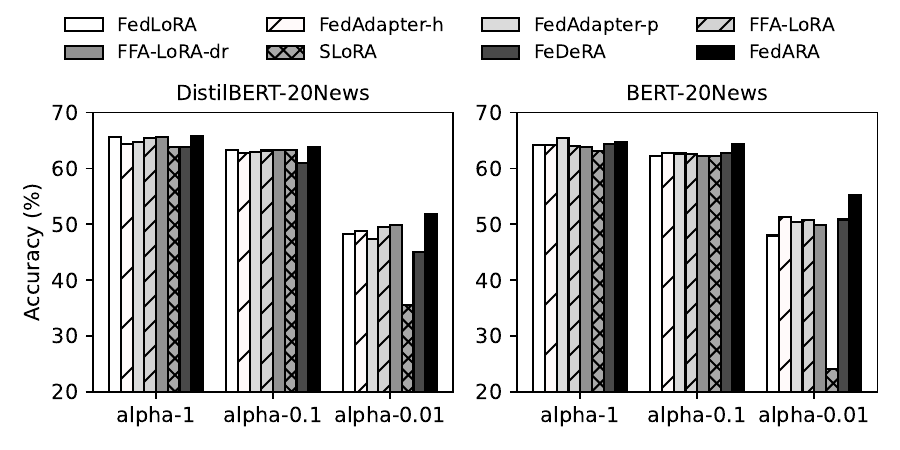}
\caption{Accuracy comparison under different data distributions, with parameter settings identical to those in Table~\ref{tab:main}.}
\label{fig7}
\end{figure}

\textbf{FedARA effectively mitigates performance degradation under non-IID settings.} FedARA consistently outperforms existing methods across four datasets and two models. Using \textit{DistilBERT}, FedARA achieves an average accuracy improvement of 8.49\% over FedLoRA and 11.58\% over FedAdapter-h, while also surpassing FFA-LoRA-dr and FFA-LoRA by 6.95\% and 7.75\%, respectively. The largest gain of 37.59\% is observed on the \textit{Semeval} dataset, which has more labels and fewer samples, whereas the smallest gain of 1.47\% occurs on \textit{AG News}, with fewer labels and more samples. Moreover, FedARA significantly outperforms SLoRA and FeDeRA, which rely on modified LoRA initialization strategies.

We further conduct experiments under varying degrees of non-IID data distributions by adjusting the Dirichlet parameter \(\alpha\), where a smaller \(\alpha\) indicates higher non-IIDness.
The results are illustrated in Figure~\ref{fig7}. As \(\alpha\) decreases, the performance of FedPEFT shows a clear downward trend, which is consistent with the observations reported in~\cite{babakniya2023slora, yan2024federa}. Moreover, FedARA consistently outperforms other FedPEFT methods across different \(\alpha\) values, with the performance gap becoming particularly significant under highly skewed non-IID settings.

Within our experimental framework, SLoRA and FeDeRA underperform FedARA on certain non-IID tasks and datasets. SLoRA requires careful hyperparameter tuning during its initial sparse fine-tuning stage, especially under varying non-IID data distributions and tasks. Its advantage tends to diminish as LoRA's rapid adaptation under the Adam optimizer. The effectiveness of FeDeRA relies on alignment between the pre-trained model and the downstream task, which becomes increasingly difficult to maintain under non-IID conditions, potentially leading to performance degradation when such alignment is weak. In contrast, FedARA leverages a scaling vector and symmetric initialization to reduce client drift and promote consistent feature representations, thereby mitigating performance degradation.

\begin{figure}[tbp]
  \centering
  \includegraphics[width=\linewidth]{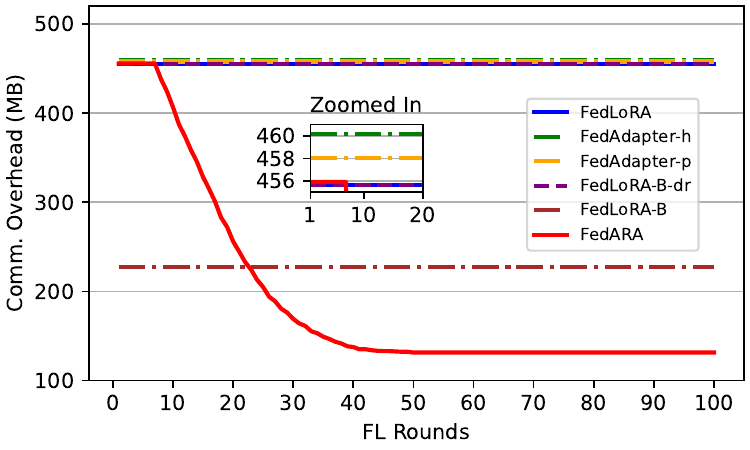}
  \caption{Comparison of communication overhead per round across baselines.}
  \label{fig8}
\end{figure}

\begin{figure}[t]
  \centering
  \includegraphics[width=\linewidth]{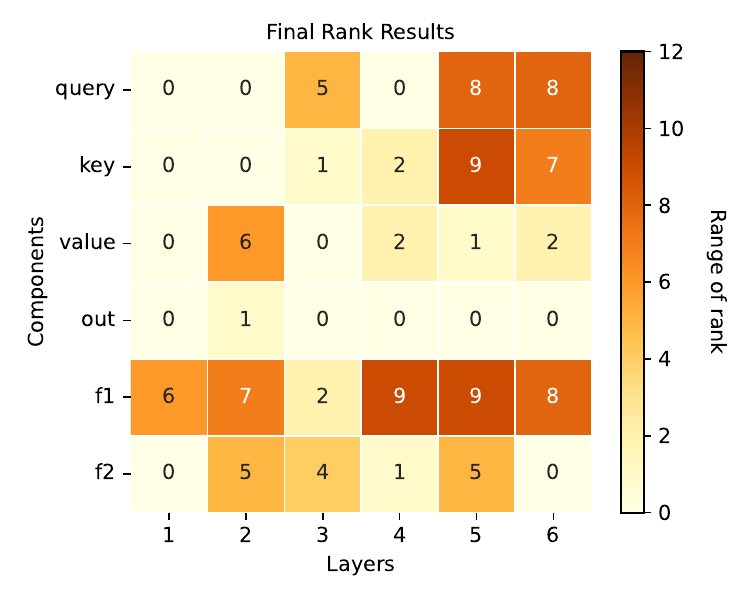}
  \caption{Results of final adaptive rank allocation  using the DistilBERT model.}
  \label{fig9}
\end{figure}

\textbf{FedARA significantly improves communication efficiency.}
FedARA reduces the communication overhead while maintaining superior performance. As shown in Figure~\ref{fig8}, FedARA begins with a similar initial communication cost (r=12) as baseline methods on the \textit{BERT} model and \textit{AG News} dataset, but quickly drops it to 131.42~MB, 71.17\% lower than its initial warm-up cost. SLoRA and FeDeRA incur the same communication cost as FedLoRA.
Table~\ref{tab:main} further shows that even when FedARA's initial rank is twice that of FedLoRA, its total communication overhead remains nearly 20\% lower than that of FedLoRA and FFA-LoRA-dr. When initial ranks are equal, FedARA reduces communication overhead by approximately 2.40\(\times\). FFA-LoRA also incurs about 20\% more overhead than FedARA. Overall, FedARA achieves the lowest communication cost across settings while consistently outperforming all baselines.

Figure~\ref{fig9} shows the final rank distribution of FedARA on the \textit{DistilBERT} model and \textit{20News} dataset. Starting from a uniform rank of 12, FedARA adaptively reduces the average rank to 3, effectively balancing performance and communication cost. Darker blocks in Figure~\ref{fig8} represent higher ranks, mainly in deeper layers and key components like \textit{query}, \textit{key}, \textit{f1}, and \textit{f2}, highlighting their role in preserving important model information, consistent with our earlier results in Figure~\ref{fig2-c}.

To evaluate the generalization and scalability of FedARA, we conduct summarization experiments using the \textit{BART} model on the \textit{CNN/DailyMail} dataset. FedARA demonstrates strong performance while achieving the best communication efficiency among all compared methods. As shown in Table~\ref{tab:sum}, FedARA improves ROUGE scores over FFA-LoRA, FFA-LoRA-dr, and FedLoRA. FedARA delivers performance comparable to SLoRA and FeDeRA while reducing communication overhead by about 60\%. These results confirm its effectiveness in generative NLP tasks with encoder-decoder architectures, in addition to classification.

\begin{table}[t]
  \caption{Results on text generation tasks fine-tuned by the BART~\cite{lewis2019bart} model for different baselines. non-IID alpha is set to 0.1. }
  \label{tab:sum}
  \centering
  \begin{tabular}{cccc}
    \toprule
    \textbf{Datasets} & \textbf{Methods} & \textbf{\makecell{Comm. \\ (GB)}} & \textbf{\makecell{ROUGE 1/2/L \\ (\%)}} \\
    \midrule
    \multirow{6}{*}{\makecell{CNN \\ /DailyMail}} 
    & FedLoRA (r=12) & 27.19 & 40.79/18.33/28.70 \\
    & SLoRA (r=12) & 27.19 & 40.90/\textbf{18.51}/28.98 \\
    & FeDeRA (r=12) & 27.19 & 40.90/18.49/\textbf{29.00} \\
    & FFA-LoRA-dr (r=24) & 27.19 & 40.71/18.28/28.66 \\
    & FFA-LoRA (r=12) & 13.60 & 40.58/18.14/28.55 \\
    & \textbf{FedARA (init r=12)} & \textbf{9.77} & \textbf{40.93}/18.49/\textbf{29.00} \\
    \bottomrule
  \end{tabular}
\end{table}

\begin{figure*}[!t]
\centering
\includegraphics[width=\linewidth]{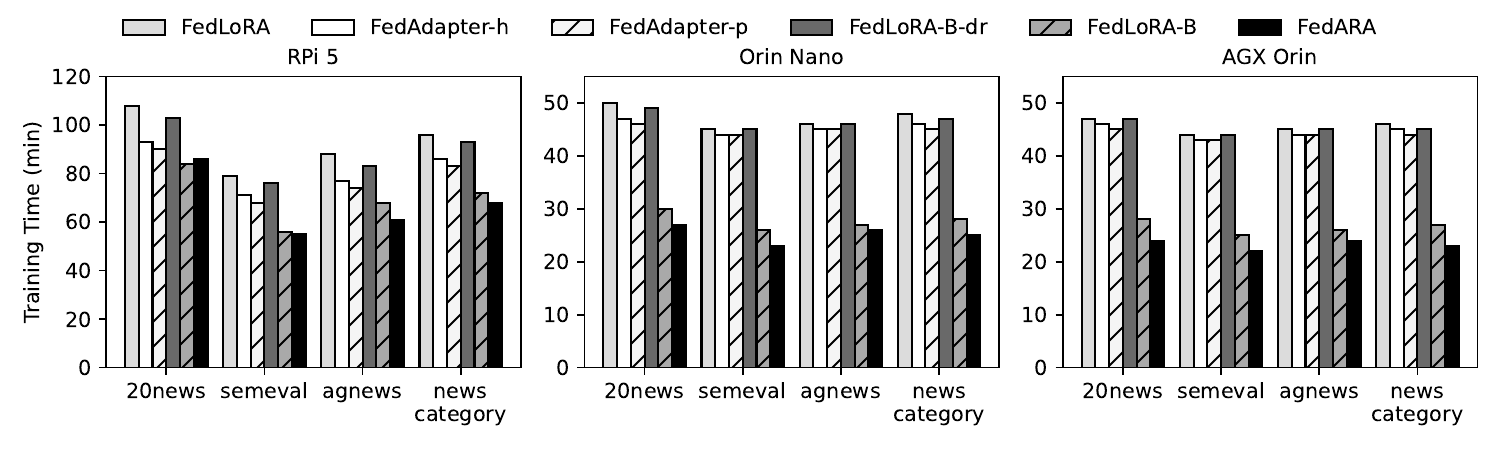}
\caption{Total training time across three different edge devices, including RPi 5, Orin Nano, and AGX Orin, each with varying hardware constraints. The local training time is measured on the \textit{DistilBERT} model to assess training time differences across these devices.}
\label{fig10}
\end{figure*}

\subsection{Training Time on Advanced Edge Devices}
We evaluate the training efficiency of FedARA and baselines on various edge devices. On Raspberry Pi 5 (RPi 5), local training time per batch (batch size = 4) is 1.00~s for \textit{DistilBERT} and 2.01~s for \textit{BERT}. AGX Orin and Orin Nano achieve significantly faster training under \textit{DistilBERT}, with 6.67\(\times\) and 5.56\(\times\) speedups over RPi 5, respectively; for \textit{BERT}, the speedups increase to 8.74\(\times\) and 6.70\(\times\). 

\begin{figure}[t]
  \centering
  \includegraphics[width=\linewidth]{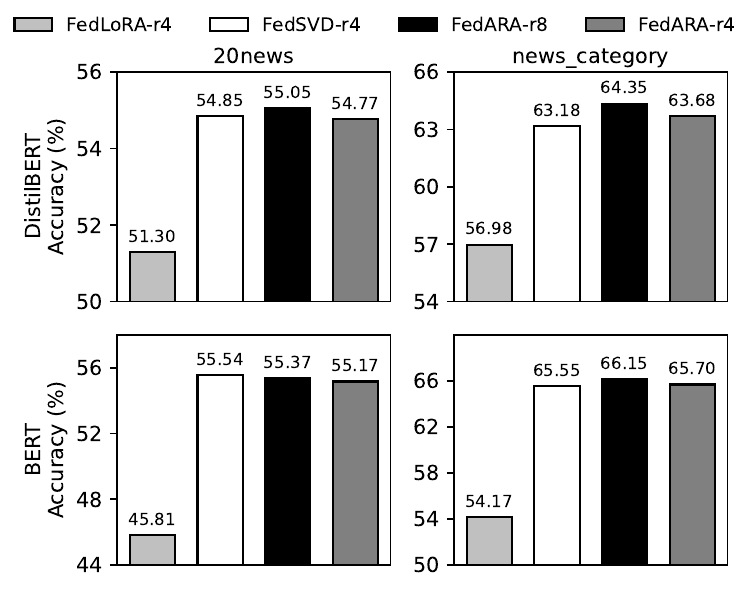}
  \caption{Effectiveness of SVD modules in mitigating non-IID problems and the performance of different initial rank configurations for FedARA.}
  \label{fig11}
\end{figure}

Figure~\ref{fig10} compares total training time for the \textit{DistilBERT} model across devices. On RPi 5, FedARA reduces training time to 68 minutes, representing 27.65\% and 24.38\% reductions over FedLoRA and FFA-LoRA-dr, respectively. On Orin Nano and AGX Orin, FedARA further lowers training time to 25 and 23 minutes, achieving up to 48.90\% reduction over FedLoRA and over 12\% compared to FFA-LoRA. These results suggest that limited computing resources on edge devices may diminish the benefits of reduced communication overhead. This observation aligns with Figure~\ref{fig2-d}, where high-end devices are communication-bound, while low-end devices remain compute-bound. We exclude comparisons with SLoRA and FeDeRA due to their noticeably higher training and computational costs compared to FedLoRA. Specifically, SLoRA incurs overhead comparable to full fine-tuning when using an optimizer masking approach~\cite{bai2024federated}, while FeDeRA introduces additional model-size-dependent costs through SVD.

\subsection{Ablation Experiments} \label{ablation}
To validate the effectiveness of our proposed methods, we conduct ablation studies comparing FedLoRA, FedARA, and FedSVD, where FedSVD replaces LoRA modules with SVD modules but lacks the adaptive rank strategy. All methods are tested with a uniform initial rank of 4 or 8.

\textbf{Ablation study of truncated SVD adaptation.} We first assess the impact of SVD modules on performance. As depicted by the first two bars of the sub-figure in Figure \ref{fig11}, FedSVD improves accuracy by an average of 7.71\% across two different models compared to FedLoRA, highlighting the importance of the SVD structure in mitigating performance degradation under non-IID settings.

\textbf{Ablation study of dynamic rank allocation.} 
To examine the role of dynamic rank allocation, we evaluate FedARA-r8 and FedARA-r4, which have initial ranks of 8 and 4, respectively. As shown in Figure \ref{fig11}, FedARA-r8 improves average performance by 0.45\% and reduces total communication overhead by nearly 20\% compared to FedSVD. Meanwhile, FedARA-r4 achieves comparable or superior performance while reducing communication overhead by approximately 2.20\(\times\). These results confirm that dynamic rank allocation maximizes communication efficiency.

\begin{figure}[t]
  \centering
  \includegraphics[width=\linewidth]{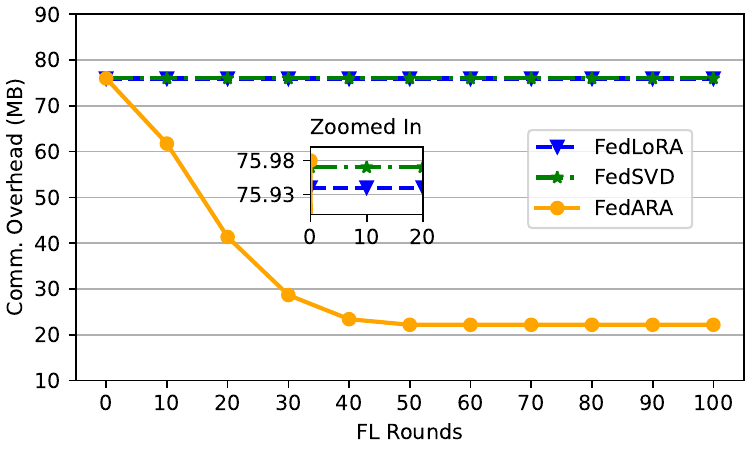}
  \caption{Effectiveness of dynamic rank allocation in improving communication efficiency per round. (Model: \textit{DistilBERT}; Dataset: \textit{20News}).}
  \label{fig12}
\end{figure}

Additionally, we assess the per-round communication overhead across FedLoRA, FedSVD, and FedARA. As illustrated in Figure \ref{fig12}, all methods begin with a similar overhead of around 75.98 MB. While FedLoRA and FedSVD maintain constant communication costs, FedARA’s dynamic rank allocation progressively reduces costs, stabilizing at 22.17 MB after 50 rounds, a 70.81\% and 70.82\% reduction compared to FedLoRA and FedSVD. This suggests that FedARA's fine-tuned model requires fewer trainable parameters, offering potential benefits for minimizing storage costs of PEFT modules for various downstream tasks.

\begin{figure}[t]
  \centering
  \includegraphics[width=\linewidth]{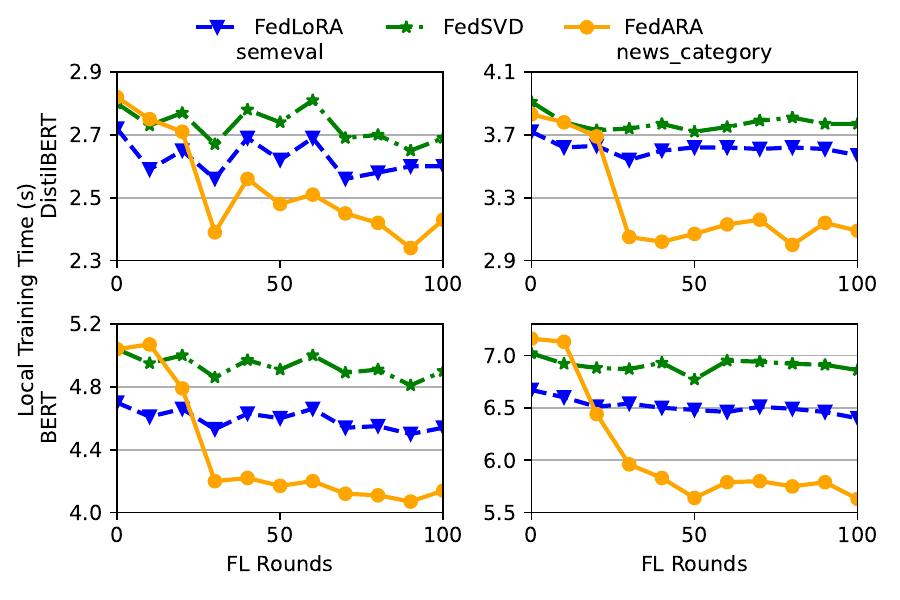}
  \vspace{-20pt} 
  \caption{Effectiveness of rank-based module pruning in reducing local training time per round.}
  \label{fig13}
\end{figure}

\begin{figure}[t]
  \centering
  \includegraphics[width=\linewidth]{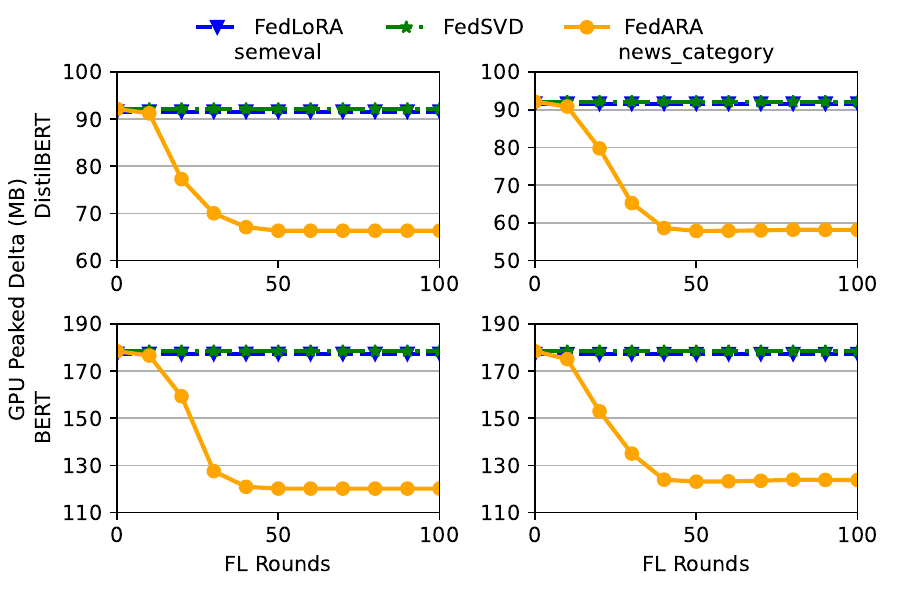}
  \vspace{-20pt} 
  \caption{Effectiveness of rank-based module pruning in lowering memory usage per round.}
  \label{fig14}
\end{figure}

\textbf{Ablation study of rank-based module pruning.} Rank-based module pruning does not impact performance, as it preserves the same parameters for communication and global aggregation. We evaluate its effect on local training time and GPU usage on \textit{Semeval} and \textit{News Category} datasets. As illustrated in Figure \ref{fig13} and \ref{fig14}, while FedARA initially exhibits a slight increase in local training time, it significantly reduces local training time over time. Compared to FedLoRA, FedARA lowers average local training time by 10.81\% and reduces GPU memory consumption by 31.67\% per round, demonstrating its efficiency in minimizing computational and storage overhead.

\subsection{Extended Experiments}
\begin{figure}[t]
  \centering
  \includegraphics[width=\linewidth]{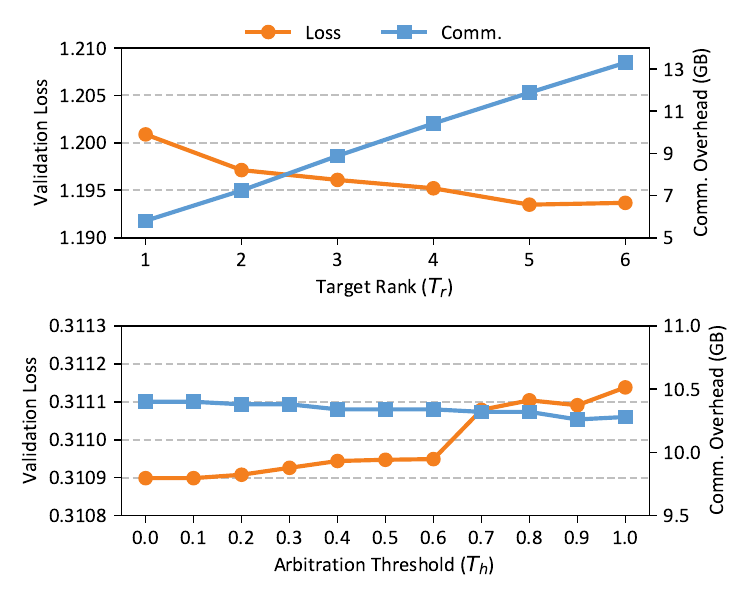}
  \vspace{-20pt} 
  \caption{Impact of target rank and arbitration threshold on validation loss and communication overhead. The upper subfigure shows results on \textit{20News}, and the lower on \textit{AG News} using the \textit{DistilBERT} model.}
  \label{fig15}
\end{figure}

To assess the impact of the target average rank (\(T_r\)) and arbitration threshold (\(T_h\)), we conduct extended experiments with an initial rank of 12 and a default threshold of 0.5. As shown in the upper part of Figure \ref{fig15}, lower target ranks reduce communication overhead but compromise validation loss, while higher ranks enhance performance at the expense of increased overhead. To balance performance and communication efficiency, we set the predefined target average rank (\(T_r\)) to one-fourth of the initial rank in this study.

Additionally, the lower part of Figure~\ref{fig15} shows that increasing the arbitration threshold (\(T_h\)) slightly increases validation loss while marginally reducing communication overhead. This is because a higher \(T_h\) enforces stricter arbitration, leading to more ranks being deemed unimportant. Notably, FedARA is relatively insensitive to \(T_h\), indicating that local rank masks tend to converge across clients. Based on these observations, we adopt 0.5 as the default value for \(T_h\) in this study.

\subsection{Local Resource Utilization}
\begin{figure}[t]
  \centering
  \includegraphics[width=\linewidth]{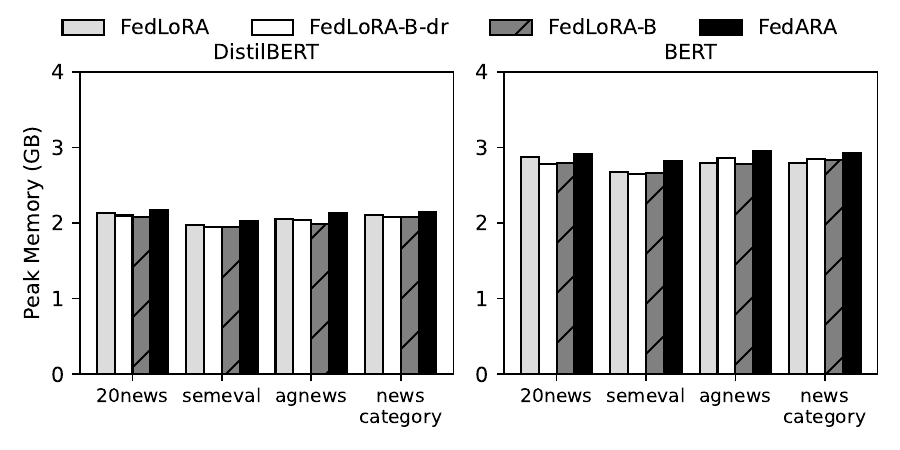}
  \vspace{-20pt} 
  \caption{Comparison of memory footprint on the Orin Nano device.}
  \label{fig16}
\end{figure}

\textbf{Memory Footprint.} Although SVD adaptation introduces an additional diagonal matrix, the memory overhead remains negligible due to the small rank \(r\) relative to the hidden layer size. As shown in Figure~\ref{fig16}, integrating SVD modules and dynamic rank allocation in FedARA leads to only a slight increase in peak memory usage compared to FedLoRA, 2.42\% for \textit{DistilBERT} and 4.40\% for \textit{BERT} on Orin Nano. Moreover, FedARA progressively reduces memory usage per round during federated fine-tuning, confirming its suitability for memory-constrained devices.

\begin{figure}[t]
  \centering
  \includegraphics[width=\linewidth]{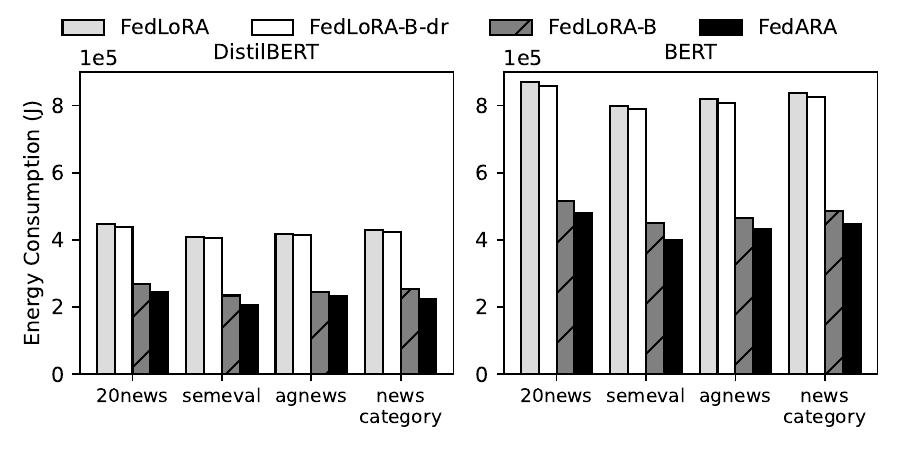}
  \vspace{-20pt} 
  \caption{Comparison of energy consumption on the Orin Nano device.}
  \label{fig17}
\end{figure}

\textbf{Energy Consumption.}  
Figure~\ref{fig17} shows the energy consumption on Orin Nano (15W) over 100 rounds with 10 clients per round. FedARA achieves the lowest energy usage, approximately \(2.3\times10^5\) J for \textit{DistilBERT} and \(4.4\times10^5\) J for \textit{BERT}, representing a 46.95\% reduction compared to FedLoRA and around 9\% compared to FFA-LoRA. These results demonstrate FedARA's ability to reduce energy consumption, thereby extending device runtime and lowering deployment costs for sustainable operation in resource-constrained environments.

\section{Related Work}
\label{relatedwork}
FedARA builds on recent advances in Parameter-Efficient Fine-Tuning (PEFT) and its extension to the federated PEFT (FedPEFT). This section reviews key developments in both areas and examines related work addressing two main challenges in FedPEFT: mitigating non-IID effects and supporting dynamic parameter configurations.

\textbf{PEFT.} PEFT enables efficient adaptation to downstream tasks by fine-tuning only a small subset of parameters, significantly reducing resource consumption compared to full fine-tuning. Based on insertion positions and integration mechanisms, PEFT methods can be categorized into: 
1) adapter-based methods~\cite{houlsby2019parameter, pfeiffer2021adapterfusion}, which insert small modules after transformer linear layers; 
2) prompt-based methods~\cite{lester2021power, li2021prefix}, which embed trainable parameters into input prompts; and 
3) LoRA-based methods~\cite{hu2022lora}, which inject low-rank matrices in parallel with transformer linear layers. 
FedARA adopts LoRA-based modules for local training, allowing seamless integration into PLMs without additional inference latency.

\textbf{FedPEFT.} FedPEFT extends PEFT to federated learning, leveraging its efficiency to reduce the substantial communication overhead of applying PEFT to PLMs. FedIT~\cite{zhang2024towards} facilitates privacy-preserving acquisition of high-quality instruction data for generative tasks. FLoRA~\cite{wang2024flora} proposes a noise-free aggregation strategy tailored to heterogeneous FL, while FlexLoRA~\cite{bai2024federated} presents an efficient framework for multi-task FL under resource heterogeneity. Other efforts focus on system-level optimization of FedPEFT~\cite{cai2023efficient, Woisetschlager2024federated} and open-source frameworks for LLMs and heterogeneous edge environments~\cite{kuang2024federatedscope, chen2023fs}. These works offer valuable context for real-world federated LLM deployment and complement our focus on efficient and adaptive federated fine-tuning under non-IID conditions.

\textbf{Non-IID Data.} Non-IID data in FL hinders model aggregation and global generalization. Traditional methods, such as client selection, optimization, and aggregation, have been effective in mitigating these effects~\cite{zou2024value, fu2023client, li2020federated, karimireddy2020scaffold, mills2023fedgbo, yao2024fedgkd, qi2023model}.  
Recent work extends this line to pre-trained models within the FedPEFT framework. SLoRA~\cite{babakniya2023slora} adopts a two-stage strategy: sparse fine-tuning for initialization, followed by LoRA adaptation. FeDeRA~\cite{yan2024federa} enhances LoRA initialization via SVD. Beyond initialization, recent studies explore asymmetric training by updating only matrix \(B\) while freezing matrix \(A\) to reduce non-IID impact~\cite{zhu2024asymmetry, zhang2023lora, sun2024improving}.  
FedARA further addresses non-IID challenges by modifying both the structure and initialization via truncated SVD adaptation, effectively slowing client drift.

\textbf{Dynamic Parameter Configurations.} Fixed-rank configurations lead to dual losses of suboptimal performance and communication inefficiency. In centralized training, AdaLoRA \cite{zhang2023adalora} introduces a sensitivity-based dynamic rank allocation method. In the context of FedPEFT, AdaFL~\cite{cai2023efficient} expands parameters via sideline trials on other clients, but introduces inference latency and additional communication costs.
SA-FedLoRA~\cite{yang2024sa} employs a two-stage simulated annealing approach: full fine-tuning with regularization to control model drift, followed by adaptive LoRA rank adjustment. However, the high cost of full fine-tuning limits its practicality on resource-constrained devices.
FedARA replaces full fine-tuning with truncated SVD adaptation for resource efficiency, and leverages rank masks to reduce communication without extra clients, enabling a more scalable solution.

\section{Conclusion}
\label{conclusion}
FedARA is a novel framework for Federated Parameter-Efficient Fine-Tuning (FedPEFT) of language models. It improves performance under non-IID data and enhances communication efficiency by integrating adaptive rank allocation with truncated SVD adaptation, while minimizing per-round local resource usage. Extensive experiments validate its superiority over existing baselines. In future work, we aim to extend FedARA to larger models and broader tasks, and explore integration with quantization to further improve adaptability on resource-constrained devices.


\bibliographystyle{IEEEtran}
\bibliography{wu-references}

\vspace{-80pt}
\begin{IEEEbiography}[{\includegraphics[width=1in,height=1.25in,clip,keepaspectratio]{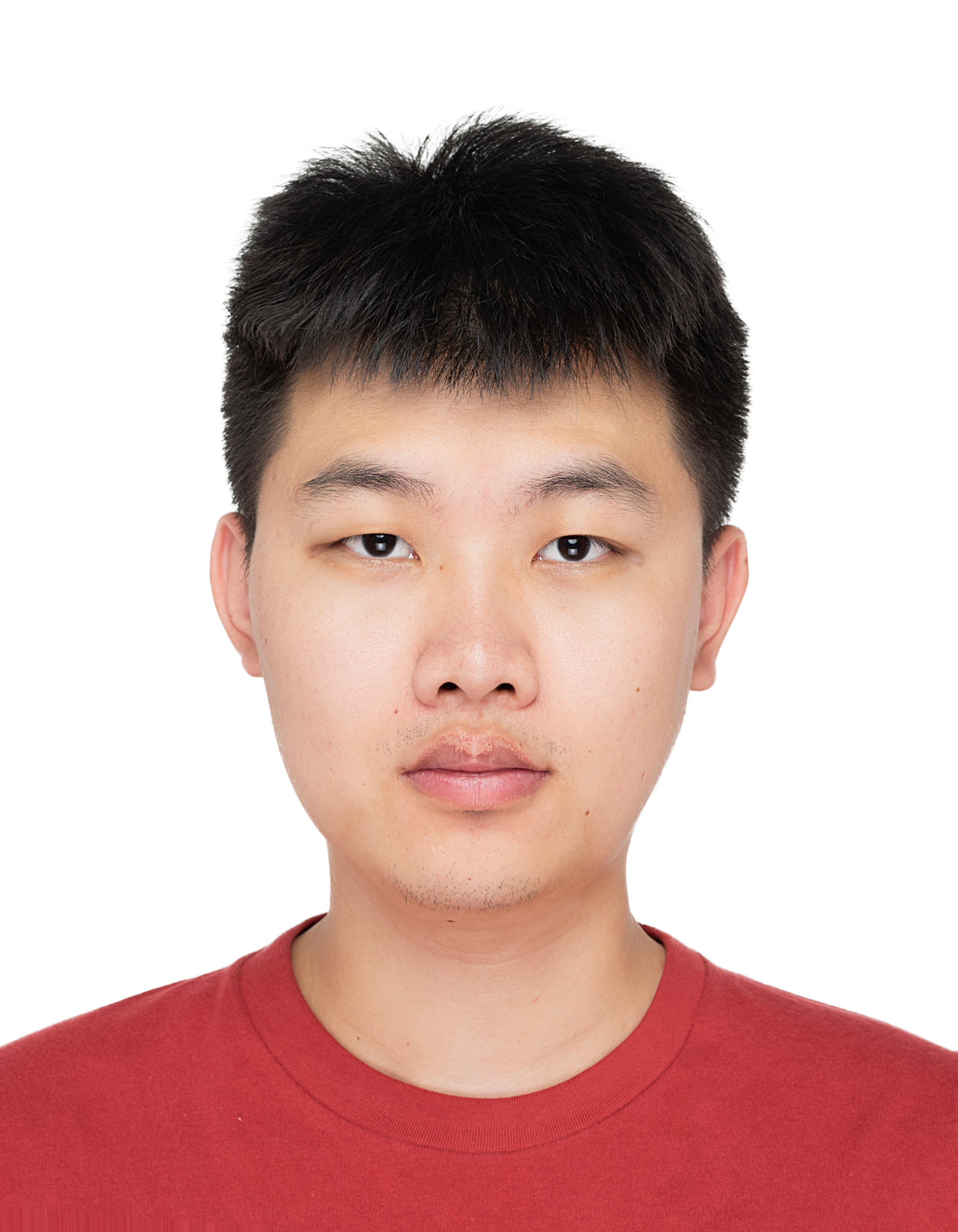}}]
{Fei Wu} received the M.Eng. degree in electronic and communication engineering from the University of Electronic Science and Technology of China (UESTC), China, in 2022. He is pursuing a Ph.D. in computer science at the University of Exeter. His research interests include efficient federated foundation models, parameter-efficient fine-tuning, edge computing, and algorithm \& hardware co-design.
\end{IEEEbiography}

\vspace{-80pt}
\begin{IEEEbiography}[{\includegraphics[width=1in,height=1.25in,clip,keepaspectratio]{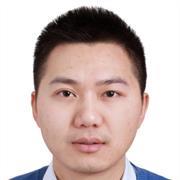}}]
{Jia Hu} received the B.Eng. and M.Eng. degrees in electronic engineering from the Huazhong University of Science and Technology (HUST), China, in 2004 and 2006, respectively, and the Ph.D. in computer science from the University of Bradford, United Kingdom, in 2010. He is an associate professor of computer science at the University of Exeter. His research interests include edge-cloud computing, resource optimization, applied machine learning, and network security.
\end{IEEEbiography}

\vspace{-80pt}
\begin{IEEEbiography}[{\includegraphics[width=1in,height=1.25in,clip,keepaspectratio]{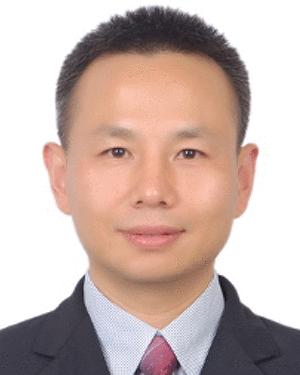}}]
{Geyong Min} received the B.Sc. degree in computer science from the Huazhong University of Science and Technology (HUST), China, in 1995, and the Ph.D. in computing science from the University of Glasgow, United Kingdom, in 2003. He is a professor of computer science at the University of Exeter. His research interests include computer networks, wireless communications, parallel and distributed computing,  ubiquitous computing, multimedia systems, modeling, and performance engineering.
\end{IEEEbiography}

\vspace{-80pt}
\begin{IEEEbiography}[{\includegraphics[width=1in,height=1.25in,clip,keepaspectratio]{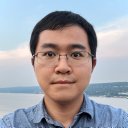}}]
{Shiqiang Wang} received the Ph.D. degree from Imperial College London, United Kingdom, in 2015. He is currently a Professor of Artificial Intelligence in the Department of Computer Science, University of Exeter, United Kingdom, after being a researcher at IBM T. J. Watson Research Center, NY, United States until Oct. 2025. His research focuses on the intersection of artificial intelligence (AI), distributed computing, and optimization. He received the IEEE Communications Society (ComSoc) Leonard G. Abraham Prize in 2021, IEEE ComSoc Best Young Professional Award in Industry in 2021, Best Paper Runner-Up of ACM MobiHoc 2025, IBM Outstanding Technical Achievement Awards (OTAA) in 2019, 2021, 2022, and 2023, multiple Invention Achievement Awards from IBM since 2016. He is an IEEE Fellow in the Class of 2026.
\end{IEEEbiography}

\end{document}